\documentclass[aps,floats,prd]{revtex}
\usepackage[latin1]{inputenc}
\usepackage{graphics}
\usepackage{subfigure}

\makeatletter

\providecommand{\LyX}{L\kern-.1667em\lower.25em\hbox{Y}\kern-.125emX\@}
\newcommand{\noun}[1]{\textsc{#1}}
\let\SF@@footnote\footnote
\def\footnote{\ifx\protect\@typeset@protect
    \expandafter\SF@@footnote
  \else
    \expandafter\SF@gobble@opt
  \fi
}
\expandafter\def\csname SF@gobble@opt \endcsname{\@ifnextchar[
  \SF@gobble@twobracket
  \@gobble
}
\edef\SF@gobble@opt{\noexpand\protect
  \expandafter\noexpand\csname SF@gobble@opt \endcsname}
\def\SF@gobble@twobracket[#1]#2{}

\usepackage[T1]{fontenc}
\usepackage[latin1]{inputenc}
\usepackage{graphics}
\usepackage{subfigure}

\makeatletter

\def\galm{\mathrm{\noun{\tiny gal}}}
\def\cmbm{\mathrm{\noun{\tiny cmb}}}

\def\comc{ CMB }
\def\Comc{ Cosmic Microwave Background }

\def\de{{\mathrm{d}}}
\def\D{\mathrm{D}}

\font\diagrm=MesDiag

\def\vertexC{\(\ \ \)\hspace{1mm}{\diagrm \char13}\(\ \ \)}
\def\vertexG{\(\ \ \)\hspace{1mm}{\diagrm \char14}\(\ \ \)}
\def\avant{\vspace{2mm}\hspace{1mm}}
\def\apres{\vspace{2mm}}
\def\mA{{\cal A}}
\def\OuvM{\(}
\def\FerM{\)}
\def\IElt{{\cal D}}
\def\Cross{{\cal X}}
\def\ii{\mathrm{i}}
\def\ee{\mathrm{e}}
\def\pabo{{\tiny \nabla }}
\def\bo{{\tiny \Delta }}
\def\Trig{{\mathcal{G}}^{\rm Ker}}
\def\CosVar{{\mathrm CosVar}}
\draft
\makeatother

\begin{document}

\twocolumn[\hsize\textwidth\columnwidth\hsize\csname @twocolumnfalse\endcsname

\title{CMB \protect\( B\protect \)-polarization to map the Large-scale Structures
of the Universe}

\author{K. Benabed, F. Bernardeau }

\address{Service de Physique Th\'eorique, C.E. de Saclay, 91191 Gif-Sur-Yvette, France}

\author{L. van Waerbeke}

\address{Canadian Institut for Theoretical Astrophysics, 60 St Georges Str., Toronto,
M5S 3H8 Ontario, Canada}

\date{\today{}}

\maketitle
\begin{abstract}
We explore the possibility of using the \( B \)-type polarization of the \Comc
to map the large-scale structures of the Universe taking advantage of the lens
effects on the CMB polarization. The functional relation between the \( B \)
component with the primordial CMB polarization and the line-of-sight mass distribution
is explicited. Noting that a sizeable fraction (at least 40\%) of the dark halo
population which is responsible of this effect can also be detected in galaxy
weak lensing survey, we present statistical quantities that should exhibit a
strong sensitivity to this overlapping. We stress that it would be a sound test
of the gravitational instability picture, independent on many systematic effects
that may hamper lensing detection in CMB or galaxy survey alone. Moreover we
estimate the intrinsic cosmic variance of the amplitude of this effect to be
less than 8\% for a \( 100\, \textrm{deg}^{2} \) survey with a \( 10' \) CMB
beam. Its measurement would then provide us with an original mean for constraining
the cosmological parameters, more particularly, as it turns out, the cosmological
constant \( \Lambda  \). 
\end{abstract}

\vspace{0.2cm} ]

\section{Introduction}

In the new era of precision cosmology we are entering in, the forthcoming experiments
will provides us with accurate data on \Comc anisotropies\cite{CMBexperiments}.
This should lead to accurate determinations of the cosmological parameters,
provided the large-scale structures of the Universe indeed formed from gravitational
instabilities of initial adiabatic scalar perturbations. It has been soon realized
however that even with the most precise experiments, the cosmological parameter
space is degenerate when the primary \comc anisotropies alone are considered\cite{BondEfst}.
Complementary data, that may be subject to more uncontrollable systematics are
thus required, such as supernovae surveys\cite{SNLmbd} (but see \cite{SNBof})
or constraints derived from the large-scale structure properties. Among the
latter, weak lensing surveys are probably the safer\cite{survlens}, but still
have not yet proved to be accurate enough with the present day observations.

Secondary \comc anisotropies (i.e. induced by a subsequent interaction of the
photons with the mass or matter fluctuations) offer opportunities for raising
this degeneracy. Lens effects\cite{TlensEffects} are particularly attractive
since they are expected to be one of the most important.They also are entirely
driven by the properties of the dark matter fluctuations, the physics of which
involve only gravitational dynamics, and are therefore totally controlled by
the cosmological parameters and not by details on galaxy or star formation rates.
More importantly an unambiguous detection of the lens effects on \comc maps
would be a precious confirmation of the gravitational instability picture. Methods
to detect the lens effects on \comc maps have been proposed recently. High order
correlation functions\cite{T4pt}, peak ellipticities\cite{peakEllip} or large
scale lens induced correlators\cite{SeljakZal} have been proposed for detecting
such effects. All of them are however very sensitive to cosmic variance since
lens effect is only a sub-dominant alteration of the \comc temperature patterns.
The situation is different when one considers the polarization properties. The
reason is that in standard cosmological models temperature fluctuations at small
scale are dominated by scalar perturbations. Therefore the pseudo-scalar part,
the so called \( B \) component, of the polarization is negligible compared
to its scalar part (the \( E \) component) and can only be significant when
\comc lens couplings are present. This mechanism has been recognized in earlier
papers\cite{B2E,BBcorde}. The aim of this paper is to study systematically
the properties of the lens induced B field and uncover its properties.

In section \ref{LensEffectSec}, we perturbatively compute the lens effect on
the \comc polarization \( E \) and \( B \) field. This first order equation
is illustrated by numerical experiments. Possibility of direct reconstruction
of the projected mass distribution is also examined. As it has already been
noted a significant fraction of the potential wells that deflect the \comc photons
can actually be mapped in local weak lensing surveys\cite{SperZut,WBBellip}.
This feature has been considered so far in relation to the \comc temperature
fluctuations. We extend in Section \ref{CrossSec} these studies to the \comc polarization
exploiting the specificities of the field found in previous section. In particular
we propose two quantities that can be built from weak lensing and \Comc polarization
surveys, the average value of which does not vanish in presence of \comc lens
effects. Compared to direct analysis of the \comc polarization, such tools have
the joint advantage of being less sensitive to systematics --systematic errors
coming from \comc mapping on one side and weak lensing measurement on the other
side have no reason to correlate!-- and so emerge even in presence of noisy
data, and of being an efficient probe of the cosmological constant. Indeed the
expected amplitude of correlation is directly sensitive to the relative length
of the optical bench, from the galaxy source plane to the \comc plane, which
is mainly sensitive to the cosmological constant. Filtering effects and cosmic
variance estimation of such quantities are considered in this section as well.

\section{Lens effects on CMB polarization}

\label{LensEffectSec}

\subsection{First order effect}

Photons emerging from the last scattering surface are deflected by the large
scale structures of the Universe that are present on the line-of-sights. Therefore
photons observed from apparent direction \( \vec{\alpha } \) must have left
the last scattering surface from a slightly different direction, \( \vec{\alpha }+\vec{\xi }(\vec{\alpha }) \),
where \( \vec{\xi } \) is the lens induced apparent displacement at that distance.
The displacement field is related to the angular gradient of the projected gravitational
potential. In the following, the lens effect will be described by the deformation
effects it induces, encoded in the amplification matrix, 
\begin{eqnarray}
\mA ^{-1} & = & \left( \begin{array}{cc}
1-\kappa -\gamma _{1} & -\gamma _{2}\\
-\gamma _{2} & 1-\kappa +\gamma _{1}
\end{array}\right) \nonumber \\
 & = & \delta _{i}^{j}+\xi ^{j}_{,i}
\end{eqnarray}
 so that 
\begin{eqnarray}
\kappa  & = & -\frac{1}{2}(\xi ^{x}_{,x}+\xi _{,y}^{y})\nonumber \\
\gamma _{1} & = & -\frac{1}{2}(\xi ^{x}_{,x}-\xi _{,y}^{y})\nonumber \\
\gamma _{2} & = & -\xi _{,x}^{y}=-\xi _{,y}^{x}.
\end{eqnarray}
 The lens effect affects the local polarization just by moving the apparent
direction of the line of sight\cite{lenspol}. Thus, if we use the Stokes parameters
\( Q \) and \( U \) to describe the local polarization vector, 
\[
\vec{P}=\left( \begin{array}{c}
Q\\
U
\end{array}\right) \]
 we can relate the observed polarization \( \hat{\vec{P}} \) to the primordial
one by the relation 
\begin{equation}
\label{QU.lens}
\hat{Q}(\vec{\alpha })=Q(\vec{\alpha }+\vec{\xi }),\qquad \hat{U}(\vec{\alpha })=U(\vec{\alpha }+\vec{\xi }).
\end{equation}
 From now on we will denote \( \hat{x} \) an observed quantity and \( x \)
the primordial one. \( \overrightarrow{\alpha }'=\vec{\alpha }+\vec{\xi } \)
is the sky coordinate system for the observer, therefore the amplification matrix
\( \mA  \) is also the Jacobian of the transformation between the source plane
and the image plane. We will restrain here our computation to the weak lensing
effect so observed quantity will not take into account any other secondary effect.
It is very important at this point to note that the lensing effect does not
produce any polarization nor rotate the Stokes parameter. In this regime its
effect reduces to a simple deformation of the polarization patterns, similar
to the temperature maps. This is the mechanism by which the geometrical properties
of the polarization field are changed.

To see that we have to consider the \emph{electric} (\( E \)) and \emph{magnetic}
(\( B \)) components instead of the Stokes parameters. At small angular scales
(we assume that a small fraction of the sky can be described by a plane), these
two quantities are defined as, 
\begin{eqnarray}
E & \equiv  & \Delta ^{-1}\left[ \left( \partial _{x}^{2}-\partial _{y}^{2}\right) \, Q+2\partial _{x}\partial _{y}\, U\right] \label{EB.Def} \\
B & \equiv  & \Delta ^{-1}\left[ \left( \partial _{x}^{2}-\partial _{y}^{2}\right) \, U-2\partial _{x}\partial _{y}\, Q\right] .\nonumber 
\end{eqnarray}
 This fields reflect non-local geometrical properties of the polarization field.
The electric component accounts for the scalar part of the polarization and
the magnetic one, the pseudo-scalar part: by parity change \( E \) is conserved,
whereas \( B \) sign is changed. As it has been pointed out in previous papers\cite{B2E,BBcorde,AllSkyHu},
lens effects partly redistribute polarization power in these two fields.

We explicit this latter effect in the weak lensing regime where distortions,
\( \kappa  \) and \( \gamma _{i} \) components are small. This is indeed expected
to be the case when lens effects by the large-scale structures are considered,
for which the typical value of the convergence field \( \kappa  \) is expected
to be \( \sim 1\% \) at 1 degree scale. The leading order effect is obtained
by simply pluging (\ref{QU.lens}) in (\ref{EB.Def}) and by expanding the result
at leading order in \( \xi  \) , \( \kappa  \), and \( \gamma  \) . Noting
that (these calculations are very similar to those done in \cite{WBBellip}),
\begin{eqnarray}
\partial _{i}\hat{X} & = & \widehat{\partial _{k}X}\cdot \left( \delta _{i}^{k}+\xi ^{k}_{,i}\right) \\
\partial _{i}\partial _{j}\hat{X} & = & \widehat{\partial _{k}\partial _{l}X}\cdot (\delta _{i}^{k}+\xi ^{k}_{,i})(\delta _{j}^{l}+\xi _{,j}^{l})\nonumber \\
 &  & \qquad +\widehat{\partial _{k}X}\cdot \xi _{,ij}^{k}\nonumber 
\end{eqnarray}
 we can write a perturbation description of the lensing effect on electric and
magnetic components of the polarization. At leading order one obtains: 
\begin{eqnarray}
\Delta \hat{E} & = & \Delta E+\xi ^{i}\partial _{i}\Delta E-2\kappa \Delta E\nonumber \\
 &  & \quad -2\delta _{ij}\left( \gamma ^{i}\Delta P^{j}+\gamma ^{i}_{,k}P^{j,k}\right) +O(\gamma ^{2})\nonumber \\
\Delta \hat{B} & = & \Delta B+\xi ^{i}\partial _{i}\Delta B-2\kappa \Delta B\label{EBLentille} \\
 &  & \quad -2\epsilon _{ij}\left( \gamma ^{i}\Delta P^{j}+\gamma ^{i}_{,k}P^{j,k}\right) +O(\gamma ^{2}),\nonumber 
\end{eqnarray}
 Where we used the fact that \( \widehat{\Delta X}=\Delta X+\xi ^{i}\partial _{i}\Delta X \)
at the leading order. The formulas for \( E \) and \( B \) are alike. The
only difference stands in the \( \delta _{ij} \) and \( \epsilon _{ij} \)
(the latter is the totally antisymmetric tensor, \( \epsilon _{11}=\epsilon _{22}=0, \)
\( \epsilon _{12}=-\epsilon _{21}=1 \)) that reflects the geometrical properties
of the two fields. The first three terms of each of these equations represent
the naive effect: the lens induced deformation of the \( E \) or \( B \) fields.
This effect is complemented by an enhancement effect (respectively \( \kappa \, \Delta E \)
and \( \kappa \, \Delta B \)) and by shear-polarization mixing terms.The latter
effects consist in two parts. One which we will call the \( \bo  \)-term that
couples the shear with second derivative of the polarization field. The other
one, hereafter the \( \pabo  \)-term, mixes gradient of the shear and polarization.
Although terms like \( \pabo  \) have been neglected in similar computations\cite{WBBellip}
we cannot do that here a priori. We will indeed show later that these two terms
have similar amplitudes.

One consequence of standard inflationary models on \comc anisotropies is the
unbalanced distribution of power between the electric (\( E \)) and magnetic
(\( B \)) component of its polarization. Adiabatic scalar fluctuations do not
induce \( B \)-type polarization and they dominate at small scales over the
tensor perturbations (namely the gravity waves). So, even though gravity waves
induce \( E \) and \( B \) type polarization in a similar amount, \emph{primary}
\comc sky is expected to be completely dominated by \( E \) type polarization
at small scales. Then for this class of models the actual magnetic component
of the polarization field is generated by the corrective part of eq. (\ref{EBLentille}),
\begin{equation}
\label{DeltaBdef}
\Delta \hat{B}=-2\epsilon _{ij}\left( \gamma ^{i}\Delta \hat{P}^{j}+\gamma ^{i}_{,k}\hat{P}^{j,k}\right) 
\end{equation}
 This result extends the direct lens effects described in Benabed \& Bernardeau\cite{BBcorde}
who focused their analysis on the lens effect due to the discontinuity of the
polarization field in case of cosmic strings. Previous studies of the weak lensing
effect on \comc showed that with lensing, the \( B \) component becomes important
at small scales\cite{B.eq.lent}. We obtain here the same result but with a
different method; eq. (\ref{DeltaBdef}) means that the polarization signal
\( P \) is redistributed by the lensing effect in a way that breaks the geometrical
properties of the primordial field. Note here that it is mathematically possible
to build a shear field that preserves these geometrical properties and that
does not create any \( B \) signal at small scales. We will discuss this problem
in Sec. \ref{KernDiscut}. It also means that \( B \) directly reflects the
properties of the shear map. We will take advantage of this feature to probe
the correlation properties of \( B \) with the projected mass distribution
in next sections.

\subsection{Lens-induced \protect\( B\protect \) maps}

We show examples of lens induced \( B \) maps. These maps have been calculated
using ``CMBSlow'' code developed by A. Riazuelo (see \cite{CMBslow}) to compute
primordial polarization maps (we use realizations of standard CDM model to illustrate
lens effects). Then various shear maps are applied. We present both true distortions,
(obtained by Delaunay triangulation used to shear the \( Q \) and \( U \)
fields), and the first order calculations given by eq. (\ref{DeltaBdef}).

\begin{figure*}
{\centering \begin{tabular}{cccc}
\resizebox*{0.24\textwidth}{!}{\includegraphics{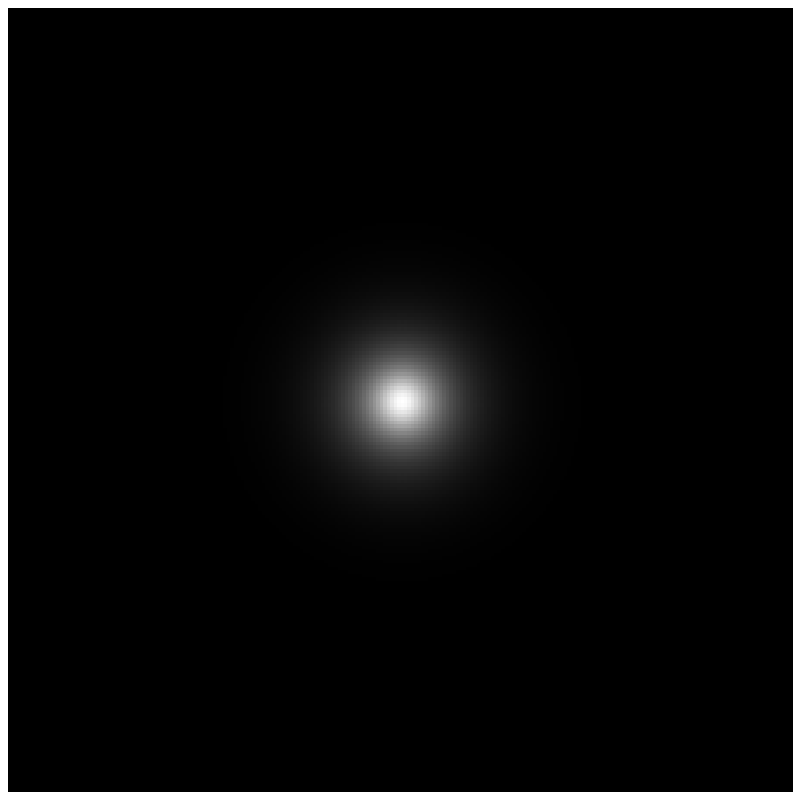}} &
\resizebox*{0.24\textwidth}{!}{\includegraphics{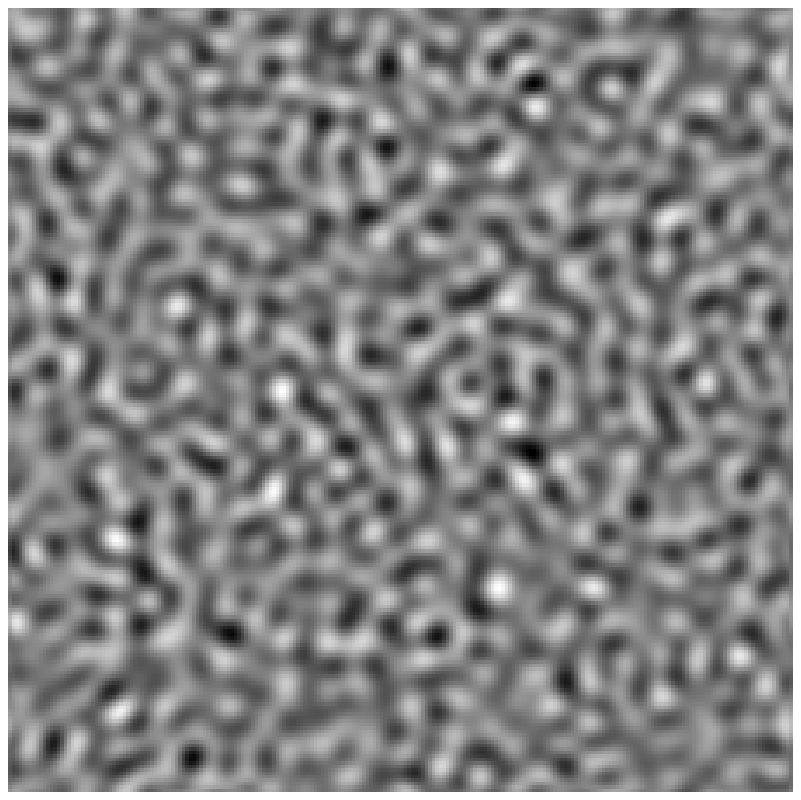}} &
\resizebox*{0.24\textwidth}{!}{\includegraphics{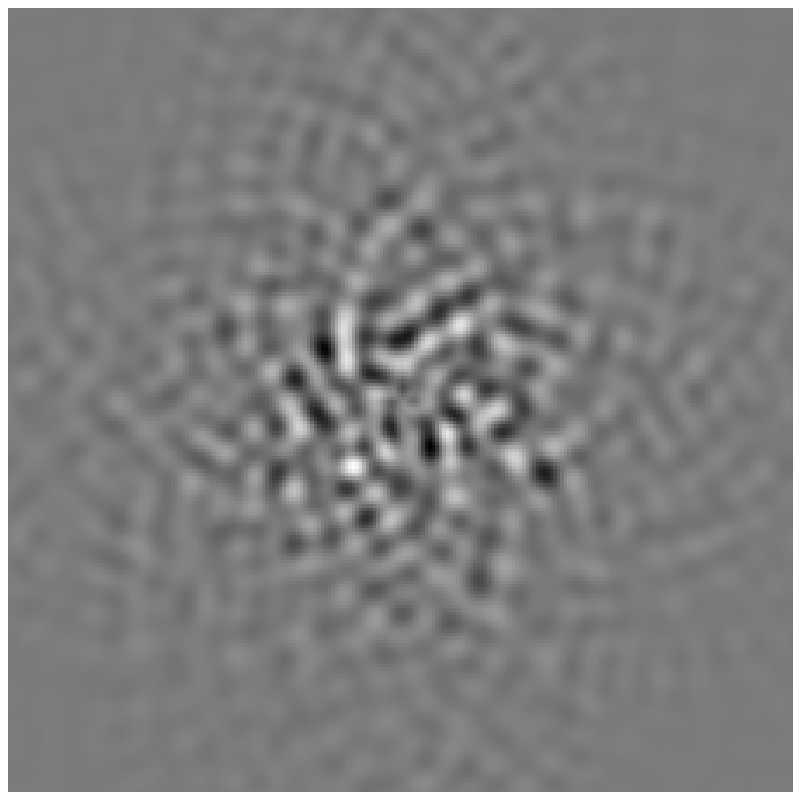}} &
\resizebox*{0.24\textwidth}{!}{\includegraphics{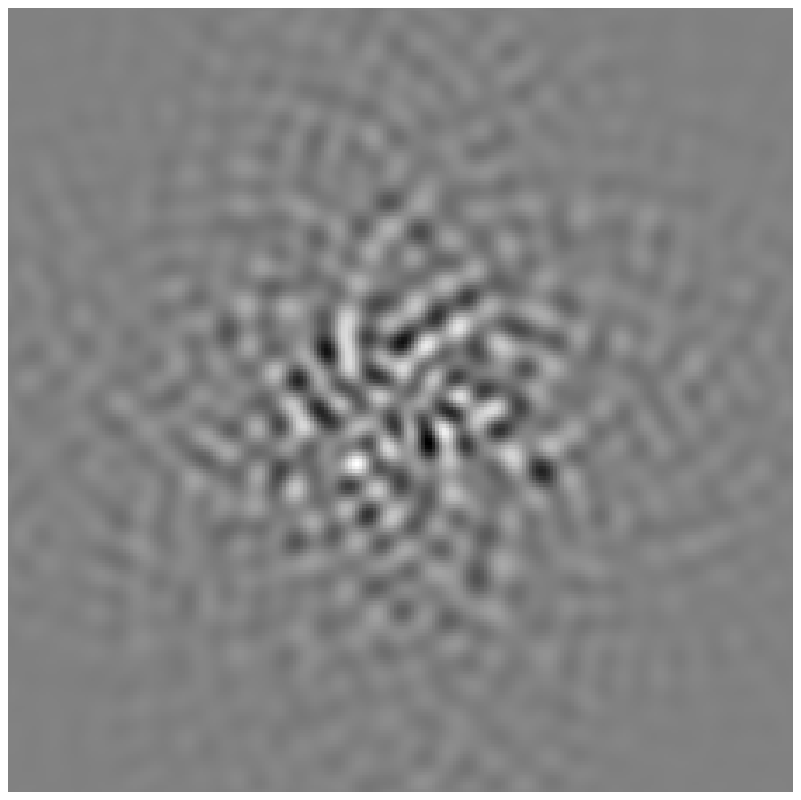}} \\
\end{tabular}\par}

\caption{\label{DelB.l.iso}Lens effect induced by a large isothermal sphere with finite
core radius. The \protect\( \kappa \protect \) map of the lens is shown on
left panel. The primordial \protect\( E\protect \) sky is presented in the
middle left panel. It has been generated for a \protect\( \Omega _{0}=0.3\protect \),
\protect\( \Lambda =0.7\protect \) model, without tensor modes. The middle
right panel displays the true reconstructed \protect\( \Delta \hat{B}\protect \)
field in a \protect\( 4.5\times 4.5\protect \) degree map and the right panel
shows the first order approximation. Note that the rosette-like shape the eye
seams to catch in \protect\( B\protect \) fields is a numerical coincidence
and has no special significance.}
\end{figure*}
 Fig. \ref{DelB.l.iso} presents the shear effect induced by an isothermal sphere
with finite core radius (and the lens edges have been suppressed by an exponential
cutoff to minimize numerical noise). The agreement between true distortion (central
panel ) and first order formula (right panel) is good. However, a close examination
of the maps reveals that some structures in the true map are slightly wider
than their counterparts in the first order map. This error is more severe in
the center, where the distortion is bigger, which is to be expected since the
limits of the validity region of first order calculations are reached.

\begin{figure*}
{\centering \begin{tabular}{ccc}
\resizebox*{0.31\textwidth}{!}{\includegraphics{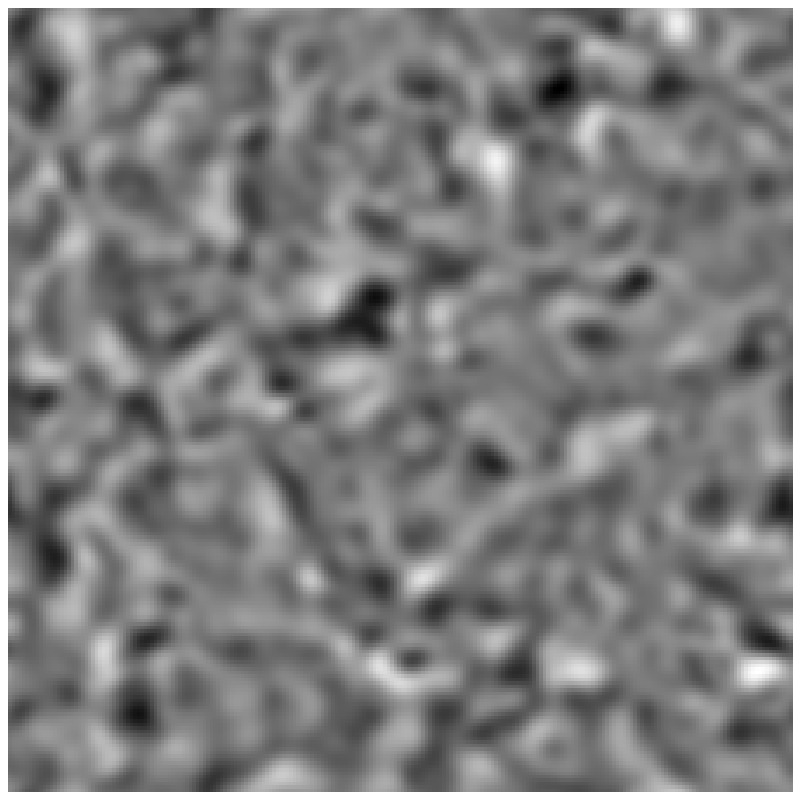}} &
\resizebox*{0.31\textwidth}{!}{\includegraphics{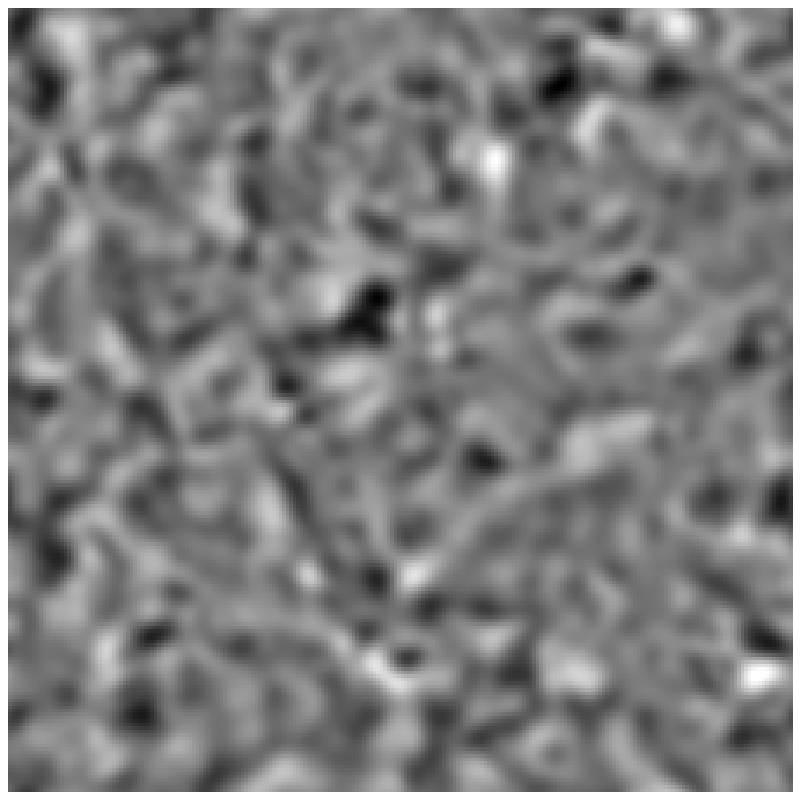}} &
\resizebox*{0.31\textwidth}{!}{\includegraphics{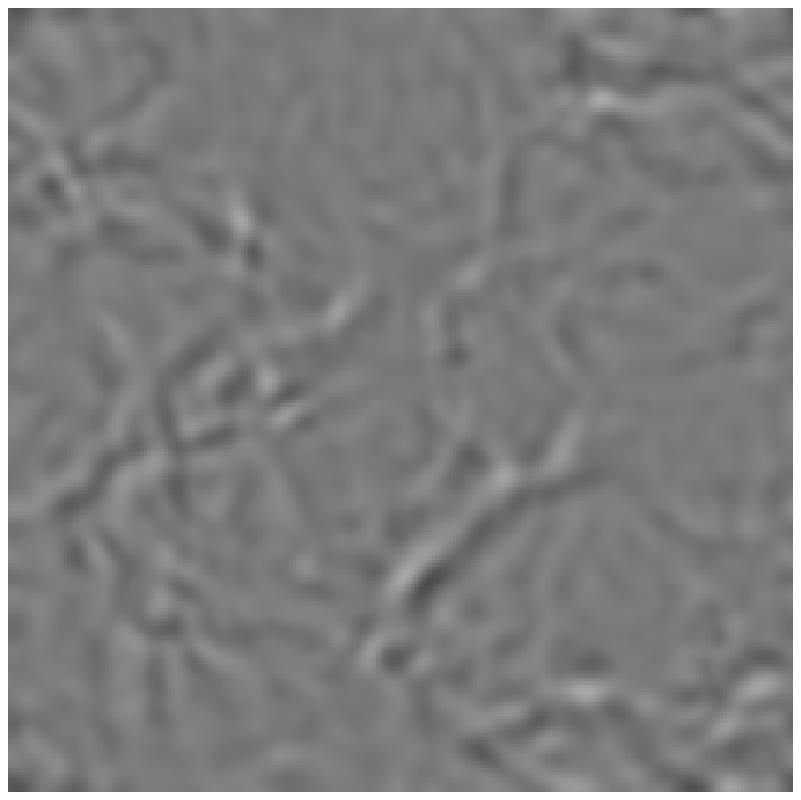}} \\
\end{tabular}\par}

\caption{The effect of a \emph{realistic} weak lensing field on B. \protect\( 2.2\times 2.2\protect \)
degree survey with \protect\( 1.8'\protect \) resolution.The left panel shows
exact distortion obtained by Delaunay triangulation. The middle one, the first
order formula result, and the right gives the difference between the two. The
three panels share the same color table. The mean amplitude in the difference
map is about 3 times smaller. }

\label{DelB.l.re}
\end{figure*}

Fig. \ref{DelB.l.re} shows the \( B \) field induced by a \emph{realistic}
distortion. We use second order Lagrangian dynamics\cite{vWBM} to create a
\( 2.5\times 2.5 \) degree map that mimics a realistic projected mass density
up to \( z=1000 \) and used its gravitational distortion to compute a typical
weak lensing-induced \( B \) map. Again we compare the \emph{exact} effect
(i.e. left panel where Delaunay triangulation is used) and the first order formula
(middle panel). Right panel shows the difference between the two maps. It reveals
the locations where the two significantly disagree. In fact most disagreements
are due to slight mismatch of the \( B \) patch positions, which lead to dipole
like effects in this map.

\begin{figure}
{\par\centering \resizebox*{0.45\textwidth}{!}{\includegraphics{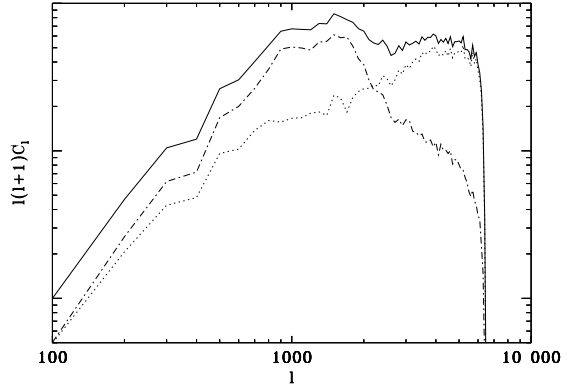}} \par}

\caption{\label{CompCl}The \protect\( C_{\ell }\protect \) of a \protect\( 100\protect \)
square degree \protect\( B\protect \) map. The solid line is the full first
order approximation formula. The dotted line gives the contribution of the \protect\( \pabo \protect \)-term.
The dash-dotted one represents the \protect\( \bo \protect \)-term. The latter
is dominant at small \protect\( \ell \protect \)s, around \protect\( \ell =1000\protect \),
that is to say for structures around \protect\( 10'\protect \). The \protect\( \pabo \protect \)
contribution gives birth to smaller structures in the \protect\( 1\sim 2\protect \)
arc-minute range.}
\end{figure}

We also show here a comparison of the two parts of the first order formula eq.
(\ref{DeltaBdef}) in order to see which of the \( \bo  \) or \( \pabo  \)
terms dominates. It would be more comfortable if one of the two terms was dominant,
however, Fig. \ref{CompCl} shows that it is not the case. Even if the \( \bo  \)-term
dominates at low (\( <1000) \) \( \ell  \), it is only twice bigger than \( \pabo  \)-one
at this scale. The inverse is true for higher (\( 3000\sim 5000) \) \( \ell  \)s.
This can be seen by looking at Fig. \ref{CompTerm} where we show the relative
amplitudes of the \( \bo  \) and \( \pabo  \) contributions. The \( \bo  \)
part gives birth to large patches (around \( 10' \)) while \( \pabo  \) panel
shows a lot more of small features.

\subsection{Direct reconstruction -- Kernel problem}

\label{KernDiscut}

The fact that the observable \( B \) is at leading order proportional to the
weak lensing signal invites us to try a direct reconstruction, similar to the
lensing mass reconstruction. In fact, we can write 
\begin{equation}
\Delta \hat{B}=-2\epsilon _{ij}\left( \gamma ^{i}\Delta \hat{P}^{j}+\gamma ^{i}_{,k}\hat{P}^{j,k}\right) \equiv {\mathrm{F}}[\gamma ]
\end{equation}
 and our reconstruction problem becomes an inversion problem for the operator
\( {\mathrm{F}} \). Unfortunately, one can prove that this problem has no unique
solution. It is due to the fact that \( \mathrm{F} \) admits a huge kernel,
in the sense that, given a polarization map, there is a wide class of shear
fields that will conserve a null \( B \) polarization. The demonstration of
this property is sketched in the following.

Since the unlensed polarization is only electric in our approximation, we can
describe it by the Laplacian of a scalar field ; 
\begin{equation}
\label{simpleE}
E\equiv \Delta \varphi \textrm{ so }\left\{ \begin{array}{l}
Q=\left( \partial _{x}^{2}-\partial _{y}^{2}\right) \, \varphi \\
U=2\, \partial _{x}\partial _{y}\, \varphi 
\end{array}.\right. 
\end{equation}
 The same holds for the shear and convergence fields 
\begin{equation}
\kappa \equiv \frac{\Delta \psi }{2},\: \gamma _{1}=\frac{1}{2}\left( \partial _{x}^{2}-\partial _{y}^{2}\right) \, \psi \textrm{ },\: \gamma _{2}=\partial _{x}\partial _{y}\, \psi .
\end{equation}
 Thus we need to know, for a given \( \varphi  \) field, whether there is any
\( \psi  \) that fulfills the equation 
\begin{equation}
\label{eqKer}
\gamma _{2}\, \Delta Q-\gamma _{1}\, \Delta U+\partial _{i}\gamma _{2}\, \partial ^{i}Q-\partial _{i}\gamma _{1}\, \partial ^{i}U=0.
\end{equation}
 \( \varphi  \) and \( \psi  \) can be written as polynomial decompositions
\begin{eqnarray}
\varphi (x,y) & = & \sum _{n,l}\, a_{nl}\, x^{n}y^{l}\nonumber \\
\psi (x,y) & = & \sum _{m,k}\, b_{mk}\, x^{m}y^{k}.\label{decPol} 
\end{eqnarray}
 Using (\ref{decPol}) in (\ref{eqKer}) we are left with a new polynomial whose
coefficients \( c_{ij} \) are sums of \( a_{nl}\times b_{mk} \) and have to
be all put to zero. With the coefficient equations in hand, it is easy to prove
that assuming all the \( b_{mk} \) coefficient up to \( m+k=N \) are known
and writing the equations \( \forall \, i+j=(N+1)-3,\, c_{ij}=0 \), we can
compute out of all the \( a_{nl} \) all but three \( b_{mk} \) with \( m+k=N+1 \).
This is somewhat similar to mass reconstruction problems from galaxy surveys
where one cannot avoid the mass sheet degeneracy. The situation is however worse
in our case since not only constant convergence but also translations and a
whole class of \( a_{nk} \) realization dependent complex deformations are
indiscernible. Thus, with the only knowledge of the \( B \) component of the
polarization one cannot, with the first order eq. (\ref{DeltaBdef}), recover
the projected mass distribution.

\section{Cross-correlating CMB maps and weak lensing surveys}

\label{CrossSec}

\subsection{Motivations}

Even with the most precise experiments it is clear that clean detections of
\( B \) component will be difficult to obtain. The magnetic polarization amplitude
induced with such a mechanism is expected to be one order of magnitude below
the electric one\cite{B.eq.lent}. Besides even if we know that there is a window
in angular scale where the other secondary effects will not interfere too much
with the detection of the lens-induced \( B \)\cite{Forg}, few is known about
removing the foregrounds\cite{bouchprun} to obtain clean maps reconstruction
algorithms would require.

These considerations lead us to look for complementary data sets to compare
\( B \) with. Although the source plane for weak lensing surveys\cite{survlens}
is much closer than for the lensed \comc fluctuations, we expect to have a significant
overlapping region in the two redshift lens distributions, so that weak lensing
surveys can map a fair fraction of the line-of-sight \comc lenses. Consequently,
weak lensing surveys can potentially provide us with shear maps correlated with
\( B \), but which have different geometrical degeneracy, noise sources and
systematics than the polarization field.

The correlation strength between the lensing effects at two different redshifts
can be evaluated. We define \( r \) as the cross-correlation coefficient between
two lens planes: 
\begin{equation}
\label{rDef}
r(z_{\galm })=\frac{\left\langle \kappa \, \kappa _{\galm }\right\rangle }{\sqrt{\left\langle \kappa ^{2}\right\rangle \left\langle \kappa _{\galm }^{2}\right\rangle }}.
\end{equation}
 In a broad range of realistic cases (see tab. \ref{Tab2r}), \( r\sim 40\% \).
To take advantage of this large overlapping we will consider quantity that cross
correlates the \comc \( B \) field and galaxy surveys. Moreover, cross-correlation
observations are expected to be insensitive to noises in weak lensing surveys
and in \comc polarization maps. This idea has already been explored for temperature
maps\cite{WBBellip}. We extend this study here taking advantage of the specific
geometrical dependences uncovered in the previous section.
\begin{table}
{\centering \begin{tabular}{ccc}
\( r \) coefficient&
\( z_{\galm }=1 \)&
\( z_{\galm }=2 \)\\
EdS, Linear&
0.42&
0.60\\
\( \Omega =0.3 \), \( \Lambda =0.7 \), Linear&
0.31&
0.50\\
\( \Omega =0.3 \), \( \Lambda =0.7 \), Non Linear&
0.40&
0.59\\
\end{tabular}\par }

\caption{ 
values of \protect\( r\protect \), the cross-correlation between two source planes (\protect\( z_{\galm} \protect \) and \protect\( z_{\cmbm} =1100\protect \)) for different models. The adopted filter scale (see Sect. \ref{FilterEffects} for details) is 2 arcmin for both weak lensing survey and \Comc observations.Non-linear \protect\( P(k)\protect \) has been computed using Peacock and Doddsmethod \protect\cite{pd96}.
}

\label{Tab2r}
\end{table}

\subsection{Definition of \protect\( b_{\bo }\protect \) and \protect\( b_{\pabo }\protect \).}

\label{btermDef}

The magnetic component of the polarization in eq. (\ref{DeltaBdef}) appears
to be built from a pure \comc part, which comes from the primordial polarization,
and a gravitational lensing part. It is natural to define \( b \), in such
a way that mimics the \( \Delta \hat{B} \) fonction dependance, by replacing
the \comc shear field by the galaxy one. 
\begin{eqnarray}
b & = & \epsilon _{ij}\left( \gamma _{\galm }^{i}\Delta \hat{P}^{j}+\gamma ^{i}_{\galm ,k}\hat{P}^{j,k}\right) \\
 & = & \epsilon _{ij}\left( \gamma _{\galm }^{i}\Delta P^{j}+\gamma _{\galm ,k}^{i}P^{j,k}\right) +O(\kappa ^{2}).\nonumber 
\end{eqnarray}
 In the following, we will label local lensing quantities, such as what one
can obtain from lensing reconstruction on galaxy surveys, with a \( \galm  \)
index. This new quantity can be viewed as a guess for the \comc polarization
\( B \) component if lensing was turned on only in a redshift range matching
the depth of galaxy surveys. The correlation coefficient of this guess with
the true \( \Delta B \) field, that is \( \left\langle \Delta \hat{B}\, b\right\rangle  \),
is expected to be quadratic both in \( P \) and in \( \gamma  \) and to be
proportional to the cross-coefficient \( r \).

For convenience, and in order to keep the objects we manipulate as simple as
possible, we will not exactly implement this scheme, as it will lead to uneven
angular derivative degrees in the two terms of resulting equations. We can,
instead, decompose the effect in the \( \bo  \) and \( \pabo  \)-part. These
two are not correlated, since their components do not share the same degrees
of angular derivation\footnote{%
generically, a random field and its derivative at the same point are not correlated. 
}. Hence, we can play the proposed game, considering the two terms of eq. (\ref{DeltaBdef})
as if they were two different fields, creating two guess-quantities that should
correlate independently with the observed \( B \) field. Following this idea
we build \( b_{\bo } \) as, 
\begin{eqnarray}
b_{\bo } & \equiv  & \epsilon _{ij}\gamma _{\galm }^{i}\Delta \widehat{P}^{j}\\
 & = & \epsilon _{ij}\gamma _{\galm }^{i}\Delta P^{j}+O(\kappa ^{2})\nonumber 
\end{eqnarray}
 which corresponds to the \( \bo  \)-term in eq. (\ref{DeltaBdef}). The amplitude
of the cross-correlation between \( \Delta B \) and \( b_{\bo } \) can easily
be estimated. At leading order, we have 
\begin{equation}
\left\langle \Delta \widehat{B}\, b_{\bo }\right\rangle =-2\epsilon _{ij}\epsilon _{kl\, }\left\langle \gamma ^{k}\gamma ^{i}_{\galm }\right\rangle \left\langle \Delta P^{l}\Delta P^{j}\right\rangle .
\end{equation}
 The corresponding \( \pabo  \) correlation is 
\begin{equation}
\left\langle \Delta \widehat{B}\, b_{\pabo }\right\rangle =-2\epsilon _{ij}\epsilon _{kl\, }\left\langle \partial _{m}\gamma ^{k}\partial _{n}\gamma ^{i}_{\galm }\right\rangle \left\langle \partial _{m}P^{l}\partial _{n}P^{j}\right\rangle 
\end{equation}
 where we have defined 
\begin{equation}
b_{\pabo }\equiv \epsilon _{ij}\partial _{k}\gamma _{\galm }^{i}\partial _{k}\widehat{P}^{j}.
\end{equation}
\begin{figure*}
{\centering \begin{tabular}{cccc}
\subfigure[  ]{\resizebox*{0.24\textwidth}{!}{\includegraphics{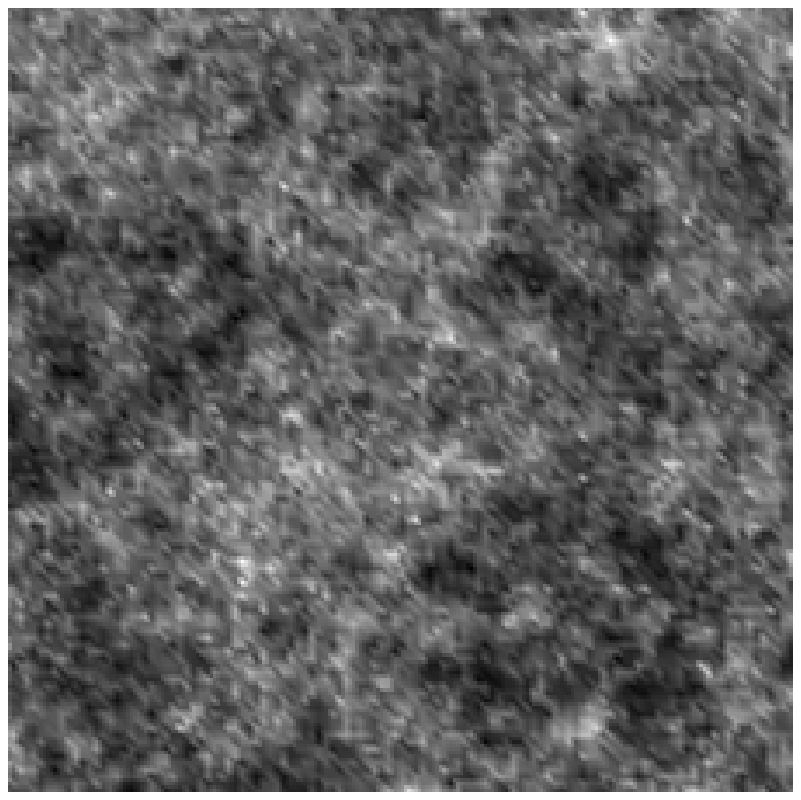}}} &
 \subfigure[  ]{\resizebox*{0.24\textwidth}{!}{\includegraphics{DelB.perttout.ps}}} &
 \subfigure[  ]{\resizebox*{0.24\textwidth}{!}{\includegraphics{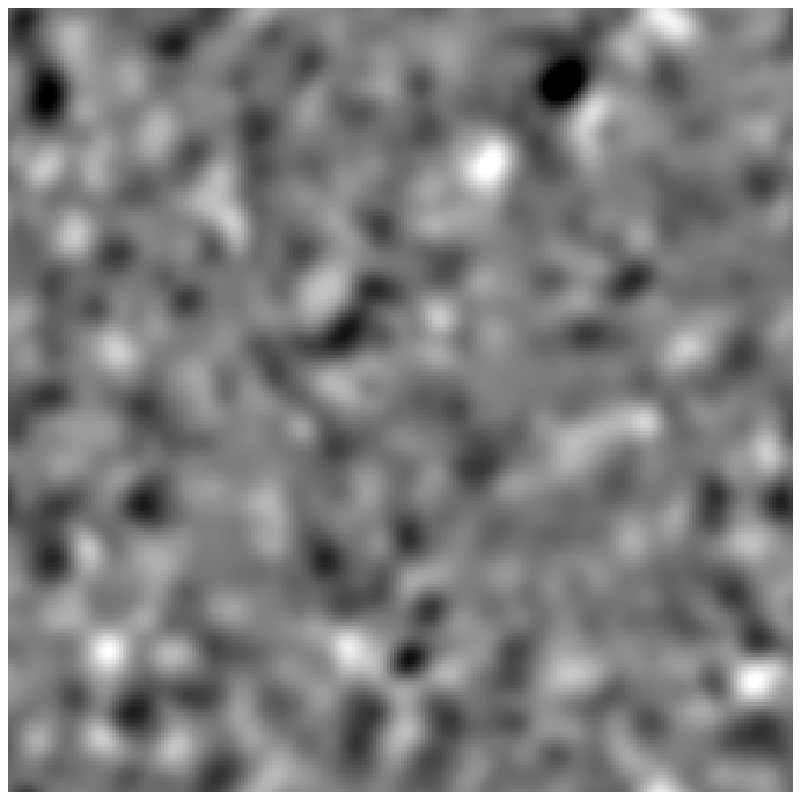}}} &
 \subfigure[  ]{\resizebox*{0.24\textwidth}{!}{\includegraphics{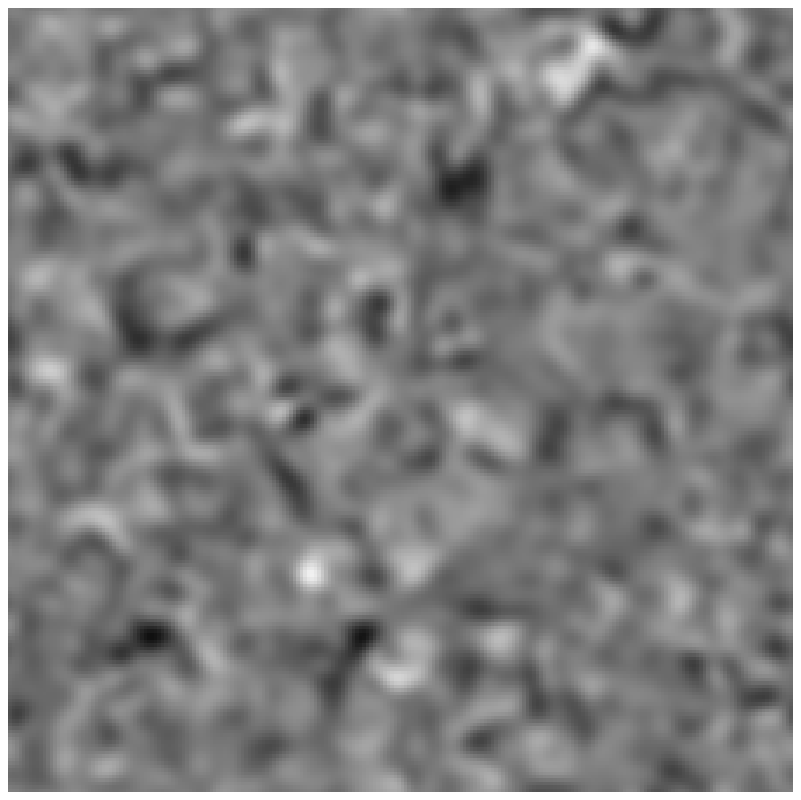}}} \\
 \subfigure[  ]{\resizebox*{0.24\textwidth}{!}{\includegraphics{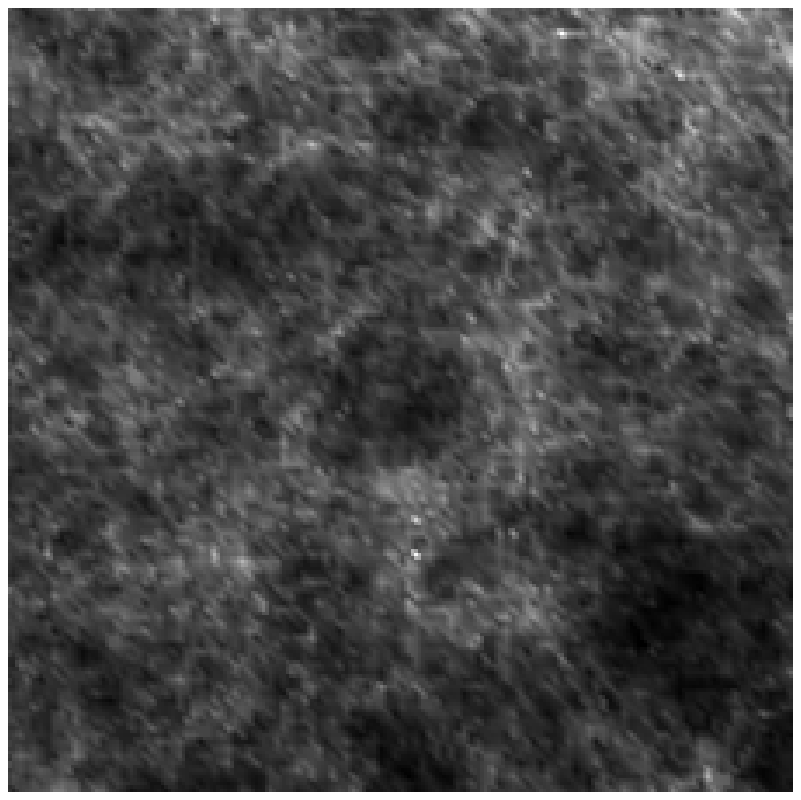}}} &
 \subfigure[  ]{\resizebox*{0.24\textwidth}{!}{\includegraphics{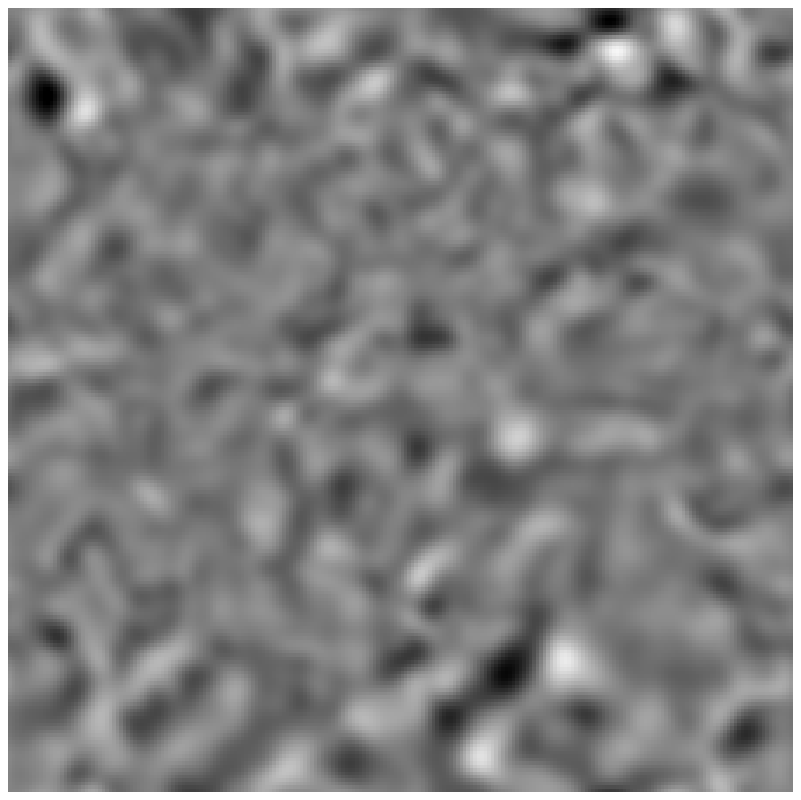}}} &
 \subfigure[  ]{\resizebox*{0.24\textwidth}{!}{\includegraphics{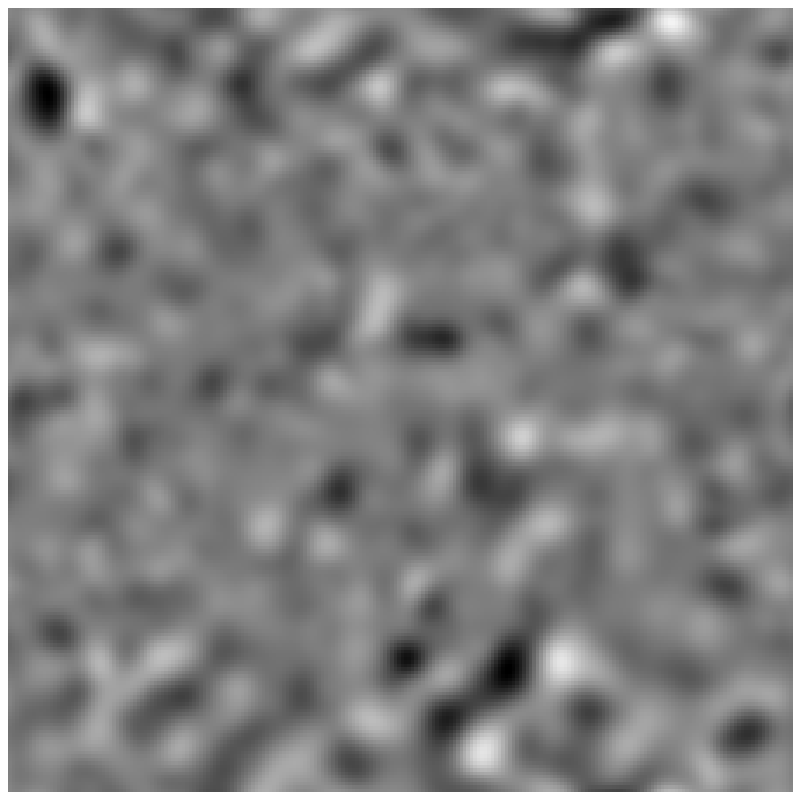}}} &
 \subfigure[  ]{\resizebox*{0.24\textwidth}{!}{\includegraphics{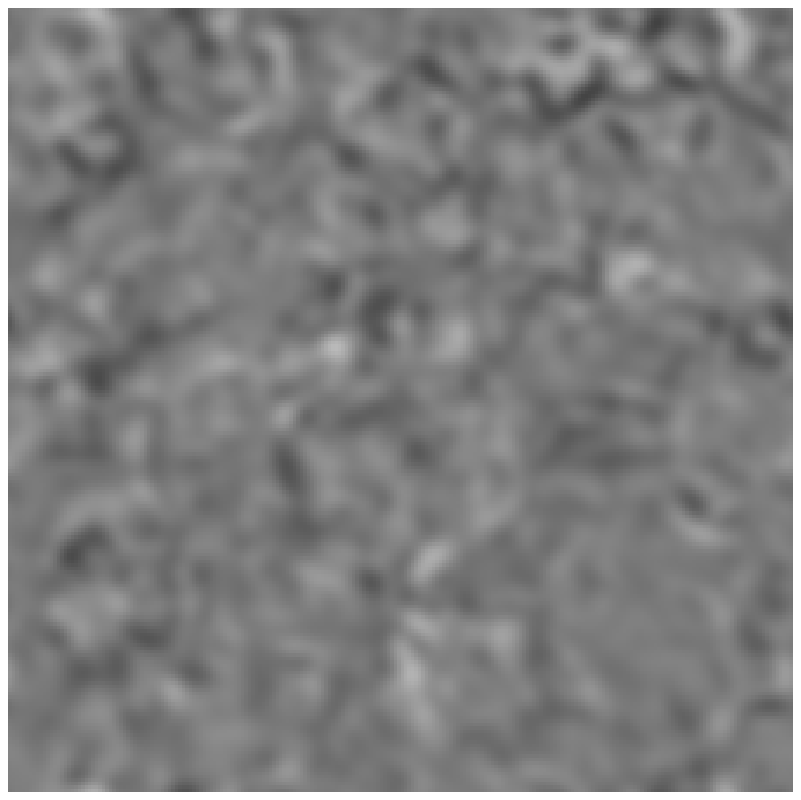}}} \\
\end{tabular}\par}

\caption{\label{CompTerm}The effect of the two terms of the perturbation formula. Top
row, the lens effect is the sum of the lenses up to recombination. Bottom row,
we use the same line -of-sight mass fluctuations but only up to redshift unity,
it represent our 'local' lensing survey. The convergence fields (left panels)
have been computed by slicing the \protect\( z\protect \)-axis and summing
up the lensing effect in each slice. Lens-lens coupling (including departure
from Born approximation) terms have been neglected, which is consistent with
our first order approximation. The convergence in each slice has been created
by using second order Lagrangian dynamics. The middle-left panels show the leading
order contribution, the middle right the \protect\( \bo \protect \) contribution
and the right the \protect\( \pabo \protect \) one. In this example, the correlation
coefficient between the two convergence maps, \protect\( r\protect \) is equal
to 0.48 at \protect\( 1.8'\protect \). The cross correlation coefficient between
the guess map (f) and the real one (b) is 0.47. It is 0.37 between the real
(b) and \protect\( \bo \protect \) (g) maps and goes down to 0.16 for the real
(b) and \protect\( \pabo \protect \) (h). }
\end{figure*}

Fig. \ref{CompTerm} shows numerical simulations presenting maps of first order
\( \Delta \widehat{B} \), its \( \bo  \) and \( \pabo  \) contributions and
the corresponding guess maps one can build with a low \( z \) shear map. The
similarities between the top maps and the bottom maps are not striking. Yet,
under close examination one can recognize individual patterns shared between
the maps. This is confirmed by the computation of the correlation coefficient
between the maps, that shows significant overlapping, between 50\% and 15\%,
depending correlation and filtering strategy. The calculations hereafter will
evaluate the theoretical correlation structure between maps given in figs. \ref{CompTerm}-b
and \ref{CompTerm}-g \& h.

For galaxy surveys, the amplification matrix is\cite{CalcKappa}, 
\begin{eqnarray}
 &  & \mA _{\galm }^{-1}(\vec{\alpha })-\textrm{Id}=-\int _{0}^{z_{\galm }}\de \chi \, w_{\galm }(\chi )\\
 &  & \qquad \times \int \frac{\de ^{3}k}{(2\pi )^{\frac{3}{2}}}\delta (\vec{k}){\mathrm{e}}^{{\mathrm{i}}\left( k_{r}\chi +\vec{k}_{\bot }D(\chi )\vec{\alpha }\right) }\nonumber \\
 &  & \qquad \times \left( \begin{array}{cc}
1+\cos (2\phi _{k_{\perp }}) & \sin (2\phi _{k_{\perp }})\\
\sin (2\phi _{k_{\perp }}) & 1-\cos (2\phi _{k_{\perp }})
\end{array}\right) \nonumber 
\end{eqnarray}
 where \( \delta (k) \) is the Fourier transform of the density contrast at
redshift \( z(\chi ) \), \( w \) is the lens efficiency function, \( \D  \)
is the angular distance, and \( \phi _{k_{\perp }} \)is the position angle
of the transverse wave-vector \( k_{\perp } \)in the \( k_{\perp }=(k_{x},\, \, k_{y}) \)
plane. Assuming a Dirac source distribution the efficiency function is given
by 
\begin{equation}
w_{\galm }(z)=\frac{3}{2}\Omega _{{\mathrm{o}}}\frac{\D _{z}\D _{z\rightarrow z_{\galm }}}{a\, \D _{z_{\galm }}}.
\end{equation}
 Note that the Fourier components \( \delta (k) \) include the density time
evolution. They are thus proportional to the growth factor in the linear theory.
The time evolution of these components is much more complicated in the nonlinear
regime (see \cite{pd96}).

Then, \( b_{\natural } \) is 
\begin{equation}
\label{bDef}
b_{\natural }(\vec{\alpha })=\int ^{\chi _{\galm }}\IElt (\chi ,\vec{l},\vec{k})\, \tilde{E}(l)\, \delta (k)\, \Trig _{\natural }\left( \vec{l},\vec{k}_{\perp }\right) 
\end{equation}
 with the integration element defined as, 
\[
\IElt (\chi ,\vec{l},\vec{k})=\de \chi \, w_{\galm }(\chi )\, \frac{\de ^{3}k}{(2\pi )^{3/2}}\frac{\de ^{2}l}{2\pi }\, \ee ^{\ii \left[ k_{r}\chi +\left( \vec{k}_{\bot }{\mathrm{D}}(\chi )+\vec{l}\right) \cdot \vec{\alpha }\right] },\]
 (it actually depends on the position of the source plane through the efficiency
function \( w(z) \)) and where \( \natural  \) stands for either \( \bo  \)
or \( \pabo  \). The geometrical kernel \( \Trig  \) is given by (using eq.
(\ref{simpleE}))
\begin{eqnarray}
\Trig _{\bo }\left( \vec{l},\vec{k}\right)  & \equiv  & l^{2}\sin 2\left( \phi _{k}-\phi _{l}\right) \\
\Trig _{\pabo }\left( \vec{l},\vec{k}\right)  & \equiv  & lk\cos \left( \phi _{k}-\phi _{l}\right) \sin 2\left( \phi _{k}-\phi _{l}\right) .
\end{eqnarray}
 This function contains all the geometrical structures of the \( \bo  \) and
\( \pabo  \) terms. We can write the same kind of equation for \( \Delta \hat{B} \).
Then, the cross-correlation is 
\begin{eqnarray}
\left\langle \Delta \hat{B}\, b_{\natural }(\vec{\alpha })\right\rangle  & = & -2\int ^{\chi _{\galm }}\IElt (\chi _{\galm },\vec{l}_{\galm },\vec{k}_{\galm })\\
 &  & \hspace {-1.5cm}\times \, \int ^{\chi _{\cmbm }}\IElt (\chi _{\cmbm },\vec{l}_{\cmbm },\vec{k}_{\cmbm })\, \, \Trig _{\natural }(\vec{l}_{\galm },\vec{k}_{\galm })\, \nonumber \\
 &  & \hspace {-1.5cm}\times \, \Trig _{\natural }(\vec{l}_{\cmbm },\vec{k}_{\cmbm })\, \left\langle \delta (\vec{k}_{\galm })\delta (\vec{k}_{\cmbm })\right\rangle \left\langle \tilde{E}(\vec{l}_{\galm })\tilde{E}(\vec{l}_{\cmbm })\right\rangle .\nonumber 
\end{eqnarray}
 The completion of this calculation requires the use of the small angle approximation,
\begin{eqnarray}
\left\langle \delta (\vec{k}_{\galm })\delta (\vec{k}_{\cmbm })\right\rangle  & = & P(k)\delta ^{3}(k_{\galm }+k_{\cmbm })\\
 &  & \hspace {-2cm}\sim P(k_{\perp })\, \delta ^{2}(k_{\galm _{\bot }}+k_{\cmbm _{\bot }})\, \delta (k_{\galm _{r}}+k_{\cmbm _{r}})\nonumber 
\end{eqnarray}
 which implies 
\begin{equation}
\vec{k}_{\galm }=-\vec{k}_{\cmbm }=\vec{k}
\end{equation}
 and after the radial components have been integrated out, 
\begin{equation}
\chi _{\galm }=\chi _{\cmbm }=\chi .
\end{equation}
 We also define the \( C_{E}(l) \) as the angular power spectrum of the \( E \)
field, 
\begin{equation}
\left\langle \tilde{E}(\vec{l}_{\galm })\tilde{E}(\vec{l}_{\cmbm })\right\rangle =C_{E}(l)\delta ^{2}(l_{\galm }-l_{\cmbm })
\end{equation}
 Eventually one gets, 
\begin{eqnarray}
\left\langle \Delta \hat{B}\, b_{\natural }(\vec{\alpha })\right\rangle  & = & -2\int ^{z_{\galm }}\! \! \! \! \de \chi \: {w}_{\galm }{w}_{\cmbm }\int \frac{\de ^{2}k\de ^{2}l}{(2\pi )^{4}}\\
 &  & \hspace {-1cm}\times \: C_{E}(l)P(k)\, \Trig _{\natural }\left( \vec{l},\vec{k}\right) ^{2}\nonumber 
\end{eqnarray}
 Then, integrating on the geometrical dependencies in \( \Trig _{\natural } \),
we have 
\begin{eqnarray}
\left\langle \Delta \hat{B}\, b_{\bo }(\vec{\alpha })\right\rangle  & = & -2\int ^{z_{\galm }}\de \chi \, {w}_{\galm }{w}_{\cmbm }\nonumber \\
 &  & \hspace {-1.5cm}\times \int \frac{\de k\de l}{2(2\pi )^{2}}\: kl^{5}C_{E}(l)P(k)\nonumber \\
 & = & -\left\langle \Delta E^{2}\right\rangle \left\langle \kappa \kappa _{\galm }\right\rangle ,
\end{eqnarray}
 and 
\begin{eqnarray}
\left\langle \Delta \hat{B}\, b_{\pabo }(\vec{\alpha })\right\rangle  & = & -\int ^{z_{\galm }}\de \chi \, {w}_{\galm }{w}_{\cmbm }\nonumber \\
 &  & \hspace {-1.5cm}\times \int \frac{\de k\de l}{2(2\pi )^{2}}\: k^{3}l^{3}C_{E}(l)P(k)\nonumber \\
 & = & -\frac{1}{2}\left\langle (\vec{\nabla }E)^{2}\right\rangle \left\langle \vec{\nabla }\kappa \cdot \vec{\nabla }\kappa _{\galm }\right\rangle ,
\end{eqnarray}
 implying that, ignoring filtering effects, we are able to measure directly
the correlation between lensing effect at \( z_{\cmbm } \) and any \( z_{\galm } \)
a weak lensing survey can access. Since \( \Delta \hat{E}=\Delta E\cdot (1+O(\kappa )) \)
we get, for the \( \bo  \) type quantity, 
\begin{eqnarray}
\left\langle \Delta \hat{E}^{2}\right\rangle  & = & \left\langle \Delta E^{2}\cdot \left( 1+O(\kappa )\right) ^{2}\right\rangle \\
 & = & \left\langle \Delta E^{2}\right\rangle \cdot \left( 1+O\left( \left\langle \kappa ^{2}\right\rangle \right) \right) .\nonumber 
\end{eqnarray}
 The same holds for \( \pabo . \) We are then able to construct two quantities
insensitive to the normalization of \comc and \( \sigma _{8} \)
\begin{eqnarray}
\Cross _{\bo } & \equiv  & \frac{\left\langle \Delta \hat{B}\, b_{\bo }(\vec{\alpha })\right\rangle }{\left\langle \Delta \hat{E}^{2}\right\rangle \left\langle \kappa _{\galm }^{2}\right\rangle }=-\frac{\left\langle \kappa \kappa _{\galm }\right\rangle }{\left\langle \kappa _{\galm }^{2}\right\rangle }\label{DefObsv1} \\
 & \sim  & -r\sqrt{\frac{\left\langle \kappa ^{2}\right\rangle }{\left\langle \kappa _{\galm }^{2}\right\rangle }}.\nonumber 
\end{eqnarray}
 and 
\begin{eqnarray}
\Cross _{\pabo } & = & \frac{\left\langle \Delta \hat{B}\, b_{\pabo }(\vec{\alpha })\right\rangle }{\left\langle (\vec{\nabla }\hat{E})^{2}\right\rangle \left\langle (\vec{\nabla }\kappa _{\galm })^{2}\right\rangle }=-\frac{1}{2}\frac{\left\langle \vec{\nabla }\kappa \cdot \vec{\nabla }\kappa _{\galm }\right\rangle }{\left\langle \vec{\nabla }\kappa _{\galm }^{2}\right\rangle }\label{DefObsv2} \\
 & \sim  & -\frac{1}{2}r_{\pabo }\sqrt{\frac{\left\langle \nabla \kappa ^{2}\right\rangle }{\left\langle \nabla \kappa _{\galm }^{2}\right\rangle }}.\nonumber 
\end{eqnarray}
 We implicitly defined \( r_{\pabo } \) like \( r \) but with \( \nabla \kappa  \)
instead of \( \kappa  \)
\begin{equation}
r_{\pabo }(z_{\galm })=\frac{\left\langle \vec{\nabla }\kappa \cdot \vec{\nabla }\kappa _{\galm }\right\rangle }{\sqrt{\left\langle (\nabla \kappa )^{2}\right\rangle \left\langle (\nabla \kappa _{\galm })^{2}\right\rangle }}.
\end{equation}
 We will see in Sect. \ref{SensCosParam} that they behave very much alike.
This result is to be compared with the formula for \( \left\langle \cos (\theta _{g})\right\rangle  \)
established in \cite{WBBellip} where the obtained quantity was going like \( r\sqrt{\left\langle \kappa ^{2}\right\rangle } \).
These calculations however have neglected the filtering effects that may significantly
affect our conclusions. These effects are investigated in next section.

\subsection{Filtering effects}

\label{FilterEffects}

In above section we conduct our calculations assuming no filtering. Obviously
we have to take it into account! We will show here that the results obtained
before hold, in certain limits, when one adds filtering effects.

In the following, we consider, for simplicity, top-hat filters only. It is expected
that other window functions will show very similar behaviors and this simplification
does not restrain the generality of our results. Let us call \( W(x) \) the
top-hat filter function in Fourier space
\begin{equation}
W(x)\equiv 2\frac{{\mathrm{J}}_{1}(x)}{x}.
\end{equation}
 \( {\mathrm{J}}_{1} \) is the first \( {\mathrm{J}} \)-Bessel function. We
will also define \( W_{i}(x) \) a general function 
\begin{equation}
W_{i}(x)\equiv 2\frac{{\mathrm{J}}_{i}(x)}{x}
\end{equation}
 where \( {\mathrm{J}}_{i} \) is the \( i^{\mathrm{th}} \) \( {\mathrm{J}} \)-Bessel
function, so that \( W=W_{1} \). Then, if \( X(\vec{\alpha }) \) is the value
of any quantity \( X \) at position \( \vec{\alpha } \) on the sky, its top-hat
filtered value can be computed as 
\begin{equation}
X_{(\theta )}(\vec{\alpha })=\int \frac{\de ^{2}k}{2\pi }\, \tilde{X}_{k}\, W(k\theta )\, \ee ^{\ii \vec{k}\cdot \vec{\alpha }},
\end{equation}
 where \( \tilde{X} \) is \( X \) Fourier transform. In the following we will
note \( X_{(\theta )} \) the filtered quantity at scale \( \theta  \).

The tricky thing for \( \left\langle \Delta \hat{B}b_{\natural }\right\rangle  \)
is that the \comc part and the low-redshift weak lensing part are \emph{a priori}
filtered at different scale. For \( \Delta \hat{B} \), which is a measured
value, its pure \comc part and its weak lensing part are filtered at the same
scale \( \theta  \). Hence, \( \hat{B} \) reads, 
\begin{eqnarray}
\Delta \hat{B}(\vec{\alpha })_{(\theta )} & = & -2\int ^{\chi _{\cmbm }}\IElt (\chi ,\vec{l},\vec{k})\, \tilde{E}(l)\, \delta (k)\nonumber \\
 &  & \hspace {-2cm}\times \left[ \Trig _{\bo }\left( \vec{l},\vec{k}_{\perp }\right) +\Trig _{\bo }\left( \vec{l},\vec{k}_{\perp }\right) \right] \, \, W\left( |\vec{k}_{\bot }{\mathrm{D}}+\vec{l}|\theta \right) 
\end{eqnarray}
 \emph{A contrario} \( b_{\natural } \) is a composite value. The \comc part
is still filtered at \( \theta  \) whereas the weak lensing part (which comes
from a weak lensing survey of galaxies) is filtered independently at another
scale which we denote \( \theta _{\galm } \). It implies that, 
\begin{eqnarray}
b_{\natural }(\vec{\alpha })_{(\theta )} & = & -2\int ^{\chi _{\galm }}\IElt (\chi ,\vec{l},\vec{k})\, \tilde{E}(l)\, \delta (k)\nonumber \\
 &  & \hspace {-1cm}\times \Trig _{\natural }\left( \vec{l},\vec{k}_{\perp }\right) \, \, W(k{\mathrm{D}}\theta _{\galm })W(l\theta ).
\end{eqnarray}

Taking filtering into account, the cross-correlation coefficient becomes, 
\begin{eqnarray}
\left\langle \Delta \hat{B}_{(\theta )}\, b_{\natural (\theta ,\theta _{\galm })}\right\rangle  & = & -2\int ^{z_{\galm }}\de \chi \, {w}_{\galm }{w}_{\cmbm }\\
 &  & \hspace {-2cm}\times \int \frac{\de ^{2}k\de ^{2}l}{(2\pi )^{4}}\: C_{E}(l)P(k)\Trig _{\natural }(\vec{l},\vec{k})\nonumber \\
 &  & \hspace {-2cm}\times W(k{\mathrm{D}}\, \theta _{\galm })W(l\theta )W(|\vec{k}{\mathrm{D}}+\vec{l}|\theta ).\nonumber 
\end{eqnarray}
 It can be shown (from the summation theorems of the Bessel functions) that,
\begin{eqnarray}
 &  & W_{1}(|\vec{k}{\mathrm{D}}+\vec{l}|\theta )=\label{WDcomp} \\
 &  & \qquad -\sum _{i=1}i\, W_{i}(k{\mathrm{D}}\theta )W_{i}(l\theta )(-\! 1)^{i}\frac{\sin i(\phi _{k}-\phi _{l})}{\sin (\phi _{k}-\phi _{l})}\nonumber 
\end{eqnarray}
 It is then possible to break the \( W\left( |\vec{k}{\mathrm{D}}+\vec{l}|\theta \right)  \)
into a sum of \( W_{i}(k{\mathrm{D}}\theta )W_{i}(l\theta ) \) with coefficients
that depend on the geometrical properties of our problem. Integrating over the
geometrical dependencies of \( \Trig _{\natural } \), leaves us with only a
few non vanishing terms in our sum, 
\begin{equation}
\label{WDecime}
\int \de \phi \, \sin ^{2}(2\phi )\frac{\sin (i\phi )}{\sin \phi }=\left\{ \begin{array}{ll}
\pi  & i\textrm{ }=\textrm{ }1\textrm{ or }i=3\\
0 & \textrm{elsewhere}
\end{array}\right. ,
\end{equation}
 for the \( \bo  \)-term and 
\begin{equation}
\label{WGDecime}
\int \de \phi \, \cos \phi \sin ^{2}(2\phi ){\sin (i\phi )}{\sin \phi }=\left\{ \begin{array}{cl}
{\pi /2} & i=1\\
{3\pi /4} & i=3\\
{\pi /4} & i=5\\
0 & \textrm{elsewhere}
\end{array}\right. ,
\end{equation}
 for the \( \pabo  \)-term. Each term can be computed exactly, and it turns
out that the terms built from \( W_{i},\, \, i>1 \) are always negligible compared
to the ones coming from \( W_{1} \). It implies that we can safely ignore the
\( W_{3} \) and \( W_{5} \) in both \( \bo  \) and \( \pabo  \) expressions,
therefore it is reasonable to assume that \( W\left( |\vec{k}{\mathrm{D}}+\vec{l}|\theta \right) =W(k{\mathrm{D}}\theta )W(l\theta ). \)
It is expected that other windows, in particular the Gaussian window function,
share similar properties. Then, taking into accounts the filtering effects,
the equations for the cross-correlations reduce to 
\begin{equation}
\left\langle \Delta \hat{B}_{(\theta )}b_{\bo (\theta ,\theta _{\galm })}\right\rangle =-\left\langle \Delta E_{(\theta )}^{2}\right\rangle \left\langle \kappa _{(\theta )}\kappa _{\galm (\theta _{\galm })}\right\rangle 
\end{equation}
 and 
\begin{equation}
\left\langle \Delta \hat{B}_{(\theta )}b_{\pabo (\theta ,\theta _{\galm })}\right\rangle =-\frac{1}{2}\left\langle \nabla E_{(\theta )}^{2}\right\rangle \left\langle \nabla \kappa _{(\theta )}\nabla \kappa _{\galm (\theta _{\galm })}\right\rangle ,
\end{equation}
 so that our correlation coefficients can be written, 
\begin{equation}
\label{DefObsvFiltre1}
\Cross _{\bo (\theta ,\theta _{\galm })}=-r_{(\theta ,\theta _{\galm })}\sqrt{\frac{\left\langle \kappa ^{2}_{(\theta )}\right\rangle }{\left\langle \kappa ^{2}_{\galm (\theta _{\galm })}\right\rangle }}
\end{equation}
 and 
\begin{equation}
\label{DefObsvFiltre2}
\Cross _{\pabo (\theta ,\theta _{\galm })}=-\frac{1}{2}r_{\pabo (\theta ,\theta _{\galm })}\sqrt{\frac{\left\langle \nabla \kappa ^{2}_{(\theta )}\right\rangle }{\left\langle \nabla \kappa ^{2}_{\galm (\theta _{\galm })}\right\rangle }}.
\end{equation}
 The results obtained in eqs. (\ref{DefObsv1}-\ref{DefObsv2}) are thus still
formally valid. Actually, eqs. (\ref{DefObsvFiltre1}-\ref{DefObsvFiltre2})
simply tell that filtering effects can simply be assumed to act independently
on the lensing effects and on the primary \Comc maps. We are left with two quantities
that only reflects the line-of-sight overlapping effects of lensing distortions.

\subsection{Sensitivity to the cosmic parameters}

\label{SensCosParam}

We quickly explore here the behavior of \( \Cross _{\natural } \) in different
sets of cosmological parameters. These quantities only depend on weak lensing
quantities. Ignoring the \( \Omega _{0} \) dependence in the angular distances
and growing factor, one would expect \( \left\langle \kappa ^{2}\right\rangle  \)
to scale like \( \Omega ^{2}_{0} \). Yet, because of the growth factor, the
convergence field exhibits a weaker sensitivity to \( \Omega _{0} \). Assuming
\( \Lambda =0 \) and a power law spectrum, we know from \cite{CalcKappa} that
\( \left\langle \kappa _{\galm }^{2}\right\rangle \propto \Omega ^{1.66}_{0} \)
for \( z_{\galm }=1 \). The same calculation leads to \( \left\langle \kappa _{\cmbm }\kappa _{\galm }\right\rangle \propto \Omega ^{1.68}_{0} \),
\( \left\langle (\nabla \kappa _{\galm })^{2}\right\rangle \propto \Omega ^{1.91}_{0} \)
and \( \left\langle \vec{\nabla }\kappa _{\cmbm }\cdot \vec{\nabla }\kappa _{\galm }\right\rangle \propto \Omega ^{1.915}_{0} \).
Then, in this limit, the quantities \( \Cross _{\natural } \) have a very low
dependence on \( \Omega _{0} \) : 
\[
\Cross _{\bo }\propto \Omega ^{0.02}_{0}\textrm{ and }\Cross _{\pabo }\propto \Omega ^{0.005}_{0}.\]
 Eventually, the \( \Cross _{\natural } \) quantities should exhibit a seizable
sensitivity to \( \Lambda  \); changing \( \Lambda  \) increases or reduces
the size of the optic bench and accordingly the overlapping between \( \kappa _{\cmbm } \)
and \( \kappa _{\galm } \).

\begin{figure}
{\par\centering \resizebox*{0.5\textwidth}{!}{\includegraphics{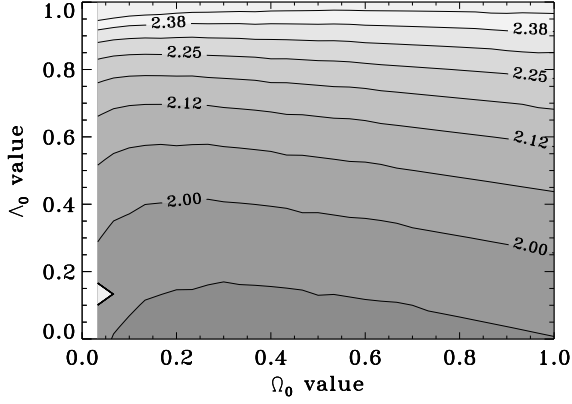}} \par}

\caption{\label{rparam}\protect\( \left\langle \kappa _{(\theta )}\kappa _{\galm (\theta _{\galm })}\right\rangle /\left\langle \kappa _{\galm (\theta _{\galm })}^{2}\right\rangle \protect \)
for a CDM model consistent with the values of (\protect\( \Omega _{0},\Lambda )\protect \).
\protect\( \theta =\theta _{\galm }=2'\protect \). }
\end{figure}

\begin{figure}
{\par\centering \resizebox*{0.5\textwidth}{!}{\includegraphics{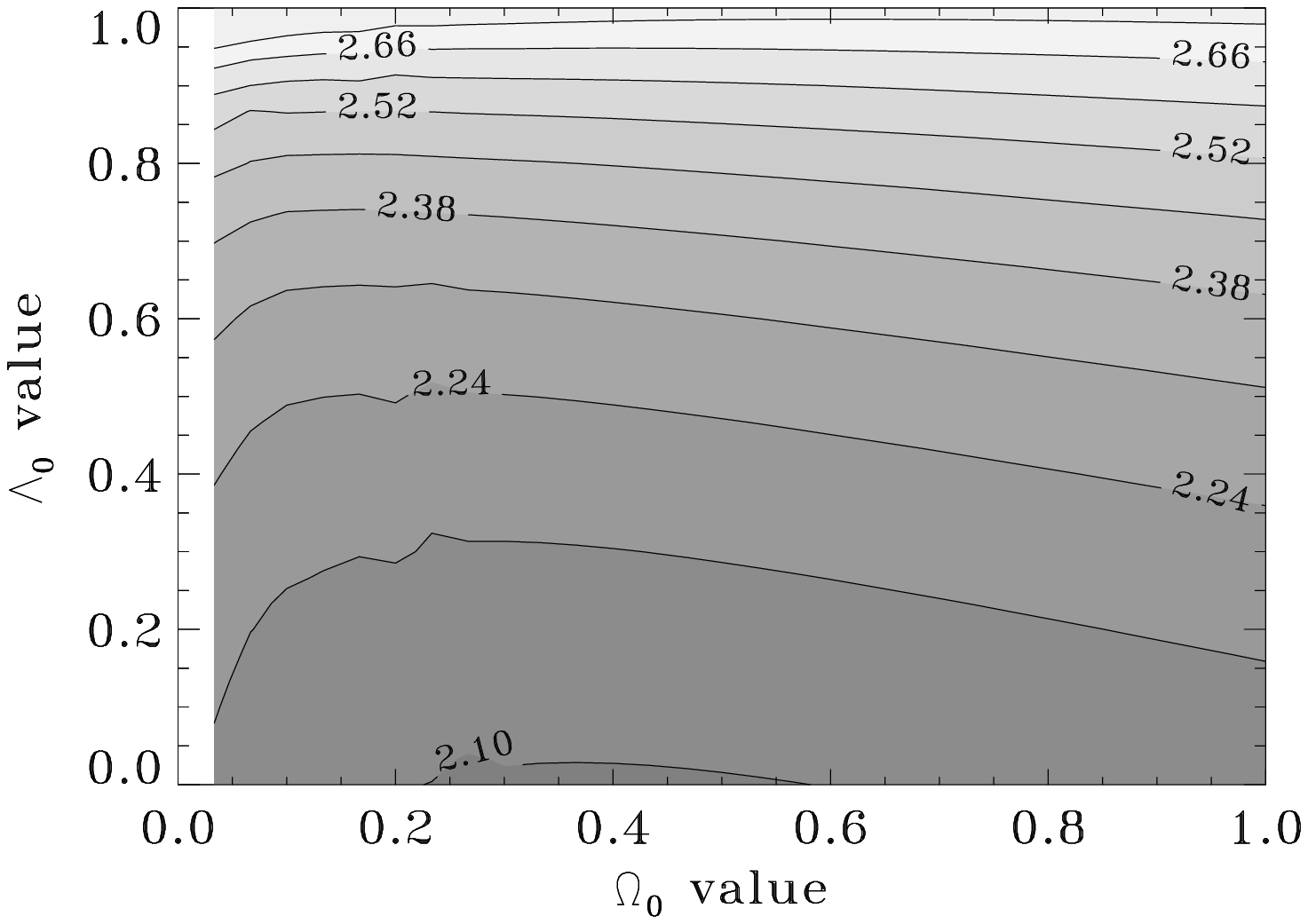}} \par}

\caption{\label{rGradparam}\protect\( \left\langle \vec{\nabla }\kappa _{(\theta )}\cdot \vec{\nabla }\kappa _{\galm (\theta _{\galm })}\right\rangle /\left\langle (\vec{\nabla }\kappa _{\galm (\theta _{\galm })})^{2}\right\rangle \protect \)
for a CDM model. \protect\( \theta =\theta _{\galm }=2'\protect \).}
\end{figure}

Figs. \ref{rparam} and \ref{rGradparam} present contour plots of the amplitude
of \( \Cross _{\bo } \) and \( \Cross _{\pabo } \) in the \( (\Omega _{0},\Lambda ) \)
plane for CDM models. They show the predicted low \( \Omega _{0} \) sensitivity
and the expected \( \Lambda  \) dependency. Both figures are very alike. This
is due to the fact that the dominant features are contained in the efficiency
function dependences on the angular distances. 

\subsection{Cosmic variance}

In previous sections we looked at the sensitivity of observable quantities which
mixed galaxy weak lensing surveys and \comc polarization detection. It is very
unlikely that both surveys will be able to cover, with a good resolution and
low foreground contamination, a large fraction of the sky. It seems however
reasonable to expect to have at our disposal patches of at least a few hundreds
square degrees. The issue we address in this section is to estimate the cosmic
variance of such a detection in joint surveys in about 100 square degrees.

The computation of cosmic variance is a classical problem in cosmological observation
\cite{Sred}. A natural estimate for an ensemble average \( \left\langle X\right\rangle  \)
is its geometrical average. If the survey has size \( \Sigma  \) then, 
\begin{equation}
\overline{X}=\frac{1}{\Sigma }\int _{\Sigma }\de ^{2}\alpha \, X(\vec{\alpha })
\end{equation}
 For a compact survey with circular shape of radius \( \Xi  \) we formally
have, 
\begin{equation}
\overline{X}=\int \frac{\de ^{2}k}{2\pi }\, \, \widetilde{X}(\vec{k})\, W(k\, \Xi ).
\end{equation}
 For sake of simplicity this is what we use in the following but we will see
that the shape of the survey has no significant consequences.

Taking \( \overline{X} \) as an estimate of \( \langle X\rangle  \) (the ensemble
average of \( X \)), leads to an error of the order \( \sqrt{\left\langle \overline{X}^{2}\right\rangle -\left\langle \overline{X}\right\rangle ^{2}} \)which
usually scales like \( 1/\sqrt{\Sigma } \) if the survey is large enough.

\begin{figure}
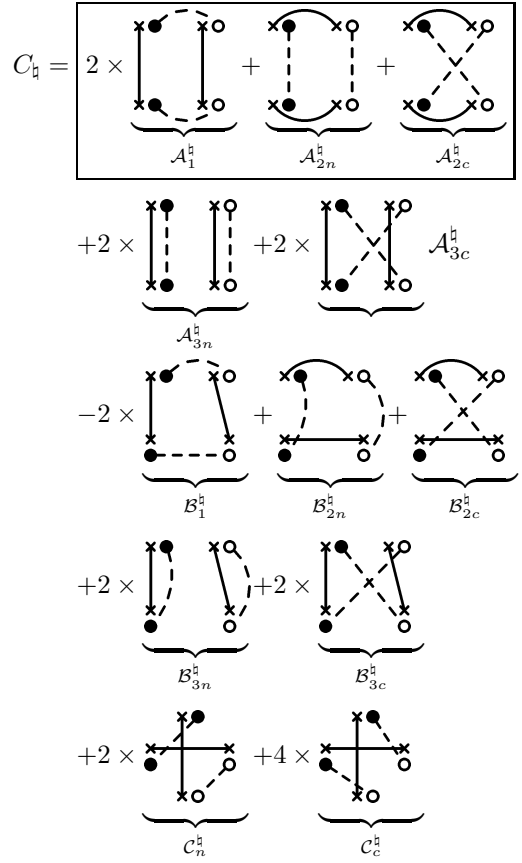

\begin{eqnarray}
C_{\natural } & = & \fbox {\OuvM 2\times \underbrace{\parbox {14mm}{\avant \diagrm \char 1\apres }}_{{\mathcal{A}}^{\natural }_{1}}+\underbrace{\parbox {14mm}{\avant \diagrm \char 2\apres }}_{{\mathcal{A}}_{2n}^{\natural }}+\underbrace{\parbox {14mm}{\avant \diagrm \char 3\apres }}_{{\mathcal{A}}^{\natural }_{2c}}\FerM }\nonumber \\
 &  & +2\times \underbrace{\parbox {14mm}{\avant \diagrm \char 4\apres }}_{{\mathcal{A}}_{3n}^{\natural }}+2\times \underbrace{\parbox {14mm}{\avant \diagrm \char 5\apres }}{\mathcal{A}}_{3c}^{\natural }\nonumber \\
 &  & -2\times \underbrace{\parbox {14mm}{\avant \diagrm \char 6\apres }}_{{\mathcal{B}}_{1}^{\natural }}+\underbrace{\parbox {14mm}{\avant \diagrm \char 7\apres }}_{{\mathcal{B}}_{2n}^{\natural }}+\underbrace{\parbox {14mm}{\avant \diagrm \char 8\apres }}_{{\mathcal{B}}_{2c}^{\natural }}\nonumber \\
 &  & +2\times \underbrace{\parbox {14mm}{\avant \diagrm \char 9\apres }}_{{\mathcal{B}}_{3n}^{\natural }}+2\times \underbrace{\parbox {14mm}{\avant \diagrm \char 10\apres }}_{{\mathcal{B}}_{3c}^{\natural }}\nonumber \\
 &  & +2\times \underbrace{\parbox {14mm}{\avant \diagrm \char 11\apres }}_{{\mathcal{C}}_{n}^{\natural }}+4\times \underbrace{\parbox {14mm}{\avant \diagrm \char 12\apres }}_{{\mathcal{C}}_{c}^{\natural }}\nonumber 
\end{eqnarray}

\caption{Diagrammatic representation of the terms contributing to the cosmic variance
of the correlation coefficients. In this representation the vertex \vertexC
represents \protect\( \Delta \hat{B}\protect \); the cross stands for the \protect\( \Delta P\protect \)
part, the dot for \protect\( \gamma _{\cmbm }\protect \). The other vertex
\vertexG represents any \protect\( b_{\natural }\protect \); the open dot stands
for \protect\( \gamma _{\galm }.\protect \) The solid lines connect \protect\( \Delta P\protect \)
terms and the dashed ones the \protect\( \gamma \protect \)-s}

\label{diagrep}
\end{figure}

When we are measuring \( \Cross _{\natural } \) on a small patch of the sky,
we are apart from the statistical value by the same kind of error. We can neglect
the errors on \( \left\langle \Delta \hat{E}^{2}\right\rangle  \), \( \left\langle (\nabla \hat{E})^{2}\right\rangle  \),
\( \left\langle (\nabla \kappa _{\galm })^{2}\right\rangle  \)and \( \left\langle \kappa _{\galm }^{2}\right\rangle  \);
those may not be the dominant source of discrepancy and can even be measured
on wider and independent samples. The biggest source of error is the measure
of \( \left\langle \Delta \hat{B}b_{\natural }\right\rangle  \). It is given
by, 
\begin{equation}
\label{ErrorDef}
C_{\natural }=\sqrt{\left\langle \left( \overline{\Delta \hat{B}b_{\natural }}-{\overline{\Delta \hat{B}}}\, {\overline{b_{\natural }}}\right) ^{2}\right\rangle -\left\langle \overline{\Delta \hat{B}b_{\natural }}-{\overline{\Delta \hat{B}}}\, {\overline{b_{\natural }}}\right\rangle ^{2}}.
\end{equation}
 The computation of (\ref{ErrorDef}) is made easier if we write explicitly
the geometrical average as a summation over \( N \) measurement points (\( N \)
can be as large as we want),
\begin{equation}
\overline{X}=\frac{1}{N}\sum ^{N}_{i=1}X(\theta _{i}),
\end{equation}
we then developed (\ref{ErrorDef}), and replace the ensemble average of the
summation sign by the geometrical average over the survey size. We are left
with a sum of correlators containing \( 8 \) fields taken at 2, 3 and 4 different
points. The calculations can be carried out analytically if we assume that all
our fields follow Gaussian statistics, which is reasonable at the scale we are
working on. In that case indeed, we can take advantage of the Wick theorem to
contract each of the \( 8 \) fields correlators in products of \( 2 \) points
correlation functions. By definition, (\ref{ErrorDef}) contains only connected
correlators, moreover the ensemble averages \( \langle \Delta \hat{B}\rangle  \)
and \( \langle b_{\natural }\rangle  \) vanish, therefore only a small fraction
of correlators among all the possible combination of the \( 8 \) fields survive.
We can use a simple diagrammatic representations to describe their geometrical
shape. All the non vanishing terms in \( C_{\natural } \) are given in Fig.
\ref{diagrep}. Each line between two vertex represents a \( 2 \) points correlation
function such as \( \langle X(\vec{\alpha }_{1})X(\vec{\alpha }_{2})\rangle  \),
and the different symbols at the vertex correspond to different \( X \) fields
(the cross stands for \( \Delta P \), the dot for \( \gamma _{\cmbm } \),
and the open dot stands for \( \gamma _{\galm } \)). The \( \mathcal{A} \)\textbf{-}terms
represent terms where the two top (and the two bottom) \( \Delta B \) and \( b_{\natural } \)
are taken at the same point, but top and bottom fields are not at the same place.
The \( \mathcal{B} \)-terms are three points diagrams: the top \( \Delta B \)
and \( b_{\natural } \) are at the same point whereas the right and left bottom
vertexes are at two different locations. The \( \mathcal{C} \) terms are four-points
diagrams, where each vertex is at a different point. To illustrate our notations,
let us write \( {\mathcal{B}}^{\natural }_{2c} \) as an example, 
\[
\begin{array}{ccc}
{\mathcal{B}}^{\natural }_{2c} & = & \langle \gamma _{\cmbm }(\vec{\alpha }_{1})\gamma _{\galm }(\vec{\alpha }_{2})\rangle \langle \gamma _{\galm }(\vec{\alpha }_{3})\gamma _{\cmbm }(\vec{\alpha }_{1})\rangle \times \\
 &  & \langle \Delta P(\vec{\alpha }_{1})\Delta P(\vec{\alpha }_{1})\rangle \langle \Delta P(\vec{\alpha }_{2})\Delta P(\vec{\alpha }_{3})\rangle 
\end{array}\]
 
We only focus on the calculation of the \( \mathcal{A} \) terms because we
can use the approximation that
\begin{equation}
{\mathcal{A}}\gg {\mathcal{B}}\gg {\mathcal{C}}.
\end{equation}
 Indeed, in perturbative theory, if the survey is large enough, the \( n \)-points
correlation functions naturally dominates over the \( n+1 \)-points correlation
function. This is true as long as the local variance is much bigger than the
autocorrelation at survey scale and we assume the surveys are still large enough
to be in this case.

The general expression for any \( \mathcal{A} \) diagram is 
\begin{eqnarray}
{\mathcal{A}}^{\natural }_{i} & = & 4\int ^{\cmbm }\IElt (\chi _{\cmbm 1},\vec{l}_{\cmbm 1},\vec{k}_{\cmbm 1})\, \IElt (\chi _{\cmbm 2},\vec{l}_{\cmbm 2},\vec{k}_{\cmbm 2})\\
 &  & \hspace {-.5cm}\times \int ^{\galm }\IElt (\chi _{\galm 1},\vec{l}_{\galm 1},\vec{k}_{\galm 1})\, \IElt (\chi _{\galm 2},\vec{l}_{\galm 2},\vec{k}_{\galm 2})\nonumber \\
 &  & \hspace {-.5cm}\times \Trig _{\natural }(\vec{l}_{\cmbm 1},\vec{k}_{\cmbm 1_{\bot }})\, \Trig _{\natural }(\vec{l}_{\cmbm 2},\vec{k}_{\cmbm 2_{\bot }})\nonumber \\
 &  & \hspace {-.5cm}\times \Trig _{\natural }(\vec{l}_{\galm 1},\vec{k}_{\galm 1_{\bot }})\, \Trig _{\natural }(\vec{l}_{\galm 2},\vec{k}_{\galm 2_{\bot }})\, {{\mathcal{M}}}_{i}\left\langle \vec{k}_{i}\mid \vec{l}_{j}\right\rangle \nonumber \\
 &  & \hspace {-.5cm}\times W(|\vec{k}_{\cmbm 1_{\perp }}{\mathrm{D}}+\vec{l}_{\cmbm 1}|\theta )\, \, W(|\vec{k}_{\cmbm 2_{\perp }}{\mathrm{D}}+\vec{l}_{\cmbm 2}|\theta )\nonumber \\
 &  & \hspace {-.5cm}\times W(k_{\galm 1_{\perp }}{\mathrm{D}_{1}}\theta _{\galm })W(l_{\galm 1}\theta )\, \, W(k_{\galm 2_{\perp }}{\mathrm{D}_{2}}\theta _{\galm })W(l_{\galm 2}\theta )\nonumber \\
 &  & \hspace {-.5cm}\times W\left( |\vec{k}_{\galm 1_{\perp }}{\mathrm{D}_{1}}+\vec{l}_{\galm 1}+\vec{k}_{\cmbm 1_{\perp }}{\mathrm{D}_{1}}+\vec{l}_{\cmbm 1}|\Xi \right) \nonumber \\
 &  & \hspace {-.5cm}\times W\left( |\vec{k}_{\galm 2_{\perp }}{\mathrm{D}_{2}}+\vec{l}_{\galm 2}+\vec{k}_{\cmbm 2_{\perp }}{\mathrm{D}_{2}}+\vec{l}_{\cmbm 2}|\Xi \right) \nonumber 
\end{eqnarray}
 where \( {\mathcal{M}}_{i} \) gives the 2-point correlations associated with
the lines of the diagram. For example : 
\begin{eqnarray}
{\mathcal{M}}_{1} & = & \! \! \left\langle \delta (\vec{k}_{\galm 1})\delta (\vec{k}_{\cmbm 1})\right\rangle \left\langle \delta (\vec{k}_{\galm 2})\delta (\vec{k}_{\cmbm 2})\right\rangle \nonumber \\
 &  & \! \! \! \! \times \left\langle \tilde{E}(l_{\galm 1})\tilde{E}(l_{\galm 2})\right\rangle \left\langle \tilde{E}(l_{\cmbm 1})\tilde{E}(l_{\cmbm 2})\right\rangle 
\end{eqnarray}

We explicit in the following the computation of \( {\mathcal{A}}^{\natural }_{1} \).
The other terms follow the same treatment or can be neglected. The lines in
the \( {\mathcal{A}}^{\natural }_{1} \) diagram give us the relations 
\begin{eqnarray}
\vec{k}_{\cmbm 1} & = & -\vec{k}_{\galm 1}=\vec{k}_{1}\nonumber \\
\vec{k}_{\cmbm 2} & = & -\vec{k}_{\galm 2}=\vec{k}_{2}\\
\vec{l}_{\cmbm 1} & = & -\vec{l}_{\cmbm 2}=\vec{l}_{\cmbm }\nonumber \\
\vec{l}_{\galm 1} & = & -\vec{l}_{\galm 2}=\vec{l}_{\galm }\nonumber 
\end{eqnarray}
 Then, using these relations and the small angular approximation, we have :
\begin{eqnarray}
{\mathcal{A}}^{\natural }_{1} & = & 4\int ^{\galm }\de \chi _{1}\de \chi _{2}\, {w}_{\cmbm 1}{w}_{\galm 1}{w}_{\cmbm 2}{w}_{\galm 2}\\
 &  & \times \int \frac{\de ^{2}k_{1}\de ^{2}k_{2}}{(2\pi )^{4}}\frac{\de ^{2}l_{\galm }\de ^{2}l_{\cmbm }}{(2\pi )^{4}}\nonumber \\
 &  & \times C_{E}(l_{\galm })C_{E}(l_{\cmbm })P(k_{1})P(k_{2})\nonumber \\
 &  & \times \Trig _{\natural }(\vec{l}_{\cmbm },\vec{k}_{1})\Trig _{\natural }(-\! \vec{l}_{\cmbm },\vec{k}_{2})\nonumber \\
 &  & \times \Trig _{\natural }(\vec{l}_{\galm },-\! \vec{k}_{1})\Trig _{\natural }(-\! \vec{l}_{\galm },-\! \vec{k}_{2})\nonumber \\
 &  & \times W(|\vec{k}_{1}{\mathrm{D}}+\vec{l}_{\cmbm }|\theta )W(|\vec{k}_{2}{\mathrm{D}}+\vec{l}_{\cmbm }|\theta )\nonumber \\
 &  & \times W(k_{1}{\mathrm{D}}\theta _{\galm })W(k_{2}{\mathrm{D}}\theta _{\galm })W^{2}(l_{\galm }\theta )\nonumber \\
 &  & \times W^{2}(|\vec{l}_{\galm }+\vec{l}_{\cmbm }|\Xi ).\nonumber 
\end{eqnarray}
 We apply the decomposition of \( W_{1}\left( |\vec{k}{\mathrm{D}}(\chi )+\vec{l}|\theta \right)  \)
we used in eq. (\ref{WDcomp}). The geometry of our problem is the same and
the result (\ref{WDecime}) still holds for the terms in \( W_{1}\left( |\vec{k}_{1}{\mathrm{D}}(\chi _{1})+\vec{l}_{\cmbm }|\theta \right)  \)
and \( W_{1}\left( |\vec{k}_{2}{\mathrm{D}}(\chi _{2})+\vec{l}_{\cmbm }|\theta \right)  \).
This however is not true for \( W_{1}^{2}\left( |\vec{l}_{\galm }+\vec{l}_{\cmbm }|\Xi \right)  \)
for which the application of the re-summation theorem does not bring any simplification.
Then, neglecting all the \( W_{3} \) parts and after integration on the \( \phi _{k_{i}} \),
for the \( \bo  \)-term, we have, 
\begin{eqnarray}
{\mathcal{A}}_{1}^{\bo } & = & \int ^{\galm }\de \chi _{1}\de \chi _{2}\, {w}_{\cmbm 1}{w}_{\galm 1}{w}_{\cmbm 2}{w}_{\galm 2}\nonumber \\
 &  & \times \int \frac{\de k_{1}\de k_{2}}{(2\pi )^{2}}\frac{\de ^{2}l_{\galm }\de ^{2}l_{\cmbm }}{(2\pi )^{4}}\: \, {l_{\galm }^{4}l_{\cmbm }^{4}}\: k_{1}k_{2}\nonumber \\
 &  & \times C_{E}(l_{\galm })C_{E}(l_{\cmbm })P(k_{1})P(k_{2})\label{A1bost1} \\
 &  & \times W^{2}(|\vec{l}_{\galm }+\vec{l}_{\cmbm }|\Xi )\cos ^{2}2\left( \phi _{l_{\cmbm }}-\phi _{l_{\galm }}\right) \nonumber \\
 &  & \times W(k_{1}{\mathrm{D}}\theta _{\galm })W(k_{2}{\mathrm{D}}\theta _{\galm })W^{2}(l_{\galm }\theta )\nonumber \\
 &  & \times W(k_{1}{\mathrm{D}}\theta )W(k_{2}{\mathrm{D}}\theta )W^{2}(l_{\cmbm }\theta ).\nonumber 
\end{eqnarray}
 Note that for the evaluation of the \( \pabo  \) part, using the same kind
of method, we obtain the same equation as eq. (\ref{A1bost1}) where \( {l_{\galm }^{4}l_{\cmbm }^{4}} \)
is replaced by \( {l_{\galm }^{2}l_{\cmbm }^{2}k_{1}^{2}k_{2}^{2}}/2 \).

\begin{figure}
{\par\centering \resizebox*{0.45\textwidth}{!}{\includegraphics{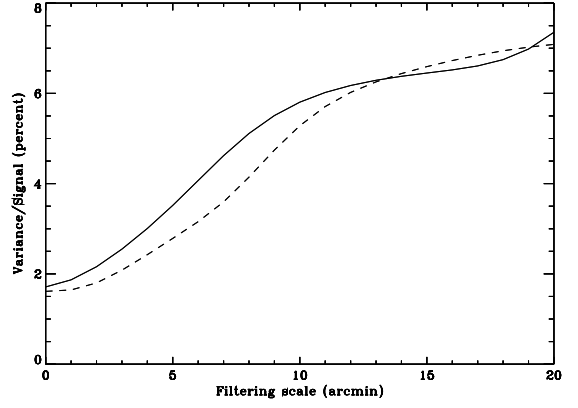}} \par}

\caption{Comparison between \protect\( \sqrt{2\, A_{1}^{\bo }/\mathrm{signal}_{\bo }}\protect \)
(solid line) and \protect\( \sqrt{2\, A_{1}^{\pabo }/\mathrm{signal}_{\pabo }}\protect \)
(dashed line). The \protect\( C_{\ell }\protect \) are from a \protect\( \Omega =0.3\protect \),
\protect\( \Lambda =0.7\protect \) model. The survey size is \protect\( 100\, \mathrm{deg}^{2}\protect \),
and Gaussian filters were used.}

\label{VarPol}
\end{figure}

We can get rid of the remaining \( W^{2}\! \left( |\vec{l}_{\galm }+\vec{l}_{\cmbm }|\Xi \right)  \)
with another approximation. The power spectrum \( C_{E}(l) \) favors large
values of \( l \) whereas \( W^{2}(|\vec{l}_{\galm }+\vec{l}_{\cmbm }|\Xi ) \)
will be non-zero for \( |\vec{l}_{\galm }+\vec{l}_{\cmbm }|\sim {1/\Xi } \).
Then for typical survey size of about one hundred square-degrees, \( |\vec{l}_{\galm }+\vec{l}_{\cmbm }|\ll l_{i} \)
and we can assume \( \vec{l}_{\galm }\sim -\vec{l}_{\cmbm } \) and \( \vec{l}_{\galm }+\vec{l}_{\cmbm }=\vec{\epsilon } \).
In this limit, \( \cos ^{2}2\left( \phi _{l_{\cmbm }}-\phi _{l_{\galm }}\right) =1 \)
and \( {\mathcal{A}}^{\natural }_{1} \) can be written 
\begin{eqnarray}
{\mathcal{A}}^{\bo }_{1} & = & \int \frac{l\, \de l}{(2\pi )^{2}}\, {l^{8}}C_{E}^{2}(l)W^{4}(l\theta )\, \int \frac{\de ^{2}\epsilon }{2\pi }\, {l^{8}}W_{1}^{2}(\epsilon \, \, \Xi )\label{A1ouf} \\
 &  & \hspace {-1cm}\times \left[ \int ^{\galm }\! \! \de \chi {w}_{\cmbm }{w}_{\galm }\int \frac{k\, \de k}{2\pi }\, P(k)\, W(k{\mathrm{D}}\theta )\, W(k{\mathrm{D}}\theta _{\galm })\right] ^{2}\nonumber 
\end{eqnarray}
 which is essentially the cosmic variance of \( \left\langle \Delta E^{2}\right\rangle  \),
for the \( \bo  \) part and of \( \left\langle (\nabla E)^{2}\right\rangle  \)
for the \( \pabo  \) one (where \( l^{8} \) in eq. (\ref{A1ouf}) is replaced
by \( l^{4}k_{1}^{2}k_{2}^{2}/2 \)). Finally we have, 
\begin{eqnarray}
\! \! \! \! \! \! \! \! \frac{{\mathcal{A}}^{\bo }_{1}}{\left\langle B_{(\theta )}b_{\bo (\theta ,\theta _{\galm })}\right\rangle ^{2}} & = & \frac{2\pi }{\Sigma }\frac{\int \de l\, l^{9}C_{E}^{2}(l)W_{1}^{4}(l\, \theta )}{\left( \int \de l\, l^{5}C_{E}(l)W_{1}^{2}(l\, \theta )\right) ^{2}}\\
 & \propto  & \textrm{Cosmic variance of }\Delta E^{2}\nonumber 
\end{eqnarray}
 where \( \Sigma =\pi \Xi ^{2} \) in case of a disc shape survey. We show in
Fig. \ref{VarPol} numerical results for a \( 100\, \mathrm{deg}^{2} \) survey
although the numerical calculations were done with a Gaussian window function
instead of a top-hat.

Numerically, for \( \theta =10' \), we get 
\begin{equation}
\label{A1Res}
\frac{{\mathcal{A}}^{\natural }_{1}}{\left\langle B_{(\theta )}b_{\natural (\theta ,\theta _{\galm })}\right\rangle ^{2}}\sim \frac{(3.7\%)^{2}}{\Sigma /100\textrm{ deg}^{2}}.
\end{equation}
We expect that for the same reasons, the \( {\mathcal{A}}^{\natural }_{2} \)
terms will be dominated by the weak lensing variance. Yet a correct evaluation
here is harder to reach. We have made this estimation within the framework of
a power law \( P(k) \). With this simplification in hand, we can write for
\( {\mathcal{A}}_{2n}^{\natural } \) (we focus only the \( \bo  \) part, but
the same discussion holds for the $ \pabo $ observable.) 

\begin{eqnarray}
\frac{{\mathcal{A}}_{2n}^{\bo }}{\left\langle B_{(\theta )}b_{\bo (\theta ,\theta _{\galm })}\right\rangle ^{2}} & = & \\
 &  & 
\hspace {-2.5cm}
\frac{1}{r^{2}}\int \de ^{2}k_{1}\de ^{2}k_{2}\, P(k_{1})P(k_{2})\cos ^{2}\left( \phi _{k_{1}}-\phi _{k_{2}}\right) \nonumber \\
 &  & 
\hspace {-1.5cm}
\times \frac{W_{1}^{2}(k_{1}\theta )W_{1}^{2}(k_{2}\theta _{\galm })W_{1}^{2}(|\vec{k}_{1}+\vec{k}_{2}|\Xi )}{\left[ \int \de ^{2}k\, P(k)W_{1}(k\theta )W_{1}(k\theta _{\galm })\right] ^{2}}.\nonumber \\
\end{eqnarray}

 The last integral behaves essentially like the cosmic variance of \( \left\langle \kappa ^{2}\right\rangle  \).
More exactly, it goes like \( 1/\sqrt{2} \) this variance. It should even be
smaller, because of the extra \( \cos ^{2} \) factor. We evaluated this cosmic
variance using the ray-tracing simulations described in \cite{jain}. These
simulations provide us with realistic convergence maps (for the cosmological
models we are interested in) with a resolution of 0.1', and a survey size of
9 square degrees. The sources have been put at a redshift unity, and the ray-lights
are propagated through a simulated Universe whose the density field has been
evolved from an initial CDM power spectrum. The measured cosmic variance of
\( \langle \kappa _{(\theta )}\kappa _{(\theta _{\galm })}\rangle  \) is about
\( 3\% \) (see Table \ref{CosVar.Tab}) when filtered at scales \( \theta _{\galm }=5' \)
and \( \theta =10' \) for a \( \Omega _{0}=0.3 \) cosmology.
\begin{table}
{\centering \begin{tabular}{ccccc}
&
\multicolumn{2}{c}{\( \CosVar \left( \left\langle \kappa ^{2}\right\rangle \right)  \) }&
\multicolumn{2}{c}{ \( \CosVar \left( \left\langle (\vec{\nabla }\kappa )^{2}\right\rangle \right)  \)}\\
&
\( \Omega _{0}=0.3 \)&
\( \Omega _{0}=1 \)&
\( \Omega _{0}=0.3 \)&
\( \Omega _{0}=1 \)\\
\( \theta =5',\, \theta _{\galm }=2.5' \) &
2.94\%&
1.86\%&
2.88\%&
2.07\%\\
\( \theta =5',\, \theta _{\galm }=5' \) &
3.02\%&
1.87\%&
2.23\%&
1.75\%\\
\( \theta =10',\, \theta _{\galm }=5' \) &
3.54\%&
2.03\%&
4.25\%&
3.02\%\\
\end{tabular}\par}

\caption{\label{CosVar.Tab}Values of the cosmic variance of \protect\( \left\langle \kappa ^{2}\right\rangle \protect \)
and \protect\( \left\langle (\vec{\nabla }\kappa )^{2}\right\rangle \protect \)
for different models and different filtering radius. The size of the survey
is \protect\( 100\, \textrm{deg}^{2}\protect \). For the \protect\( \Omega _{0}=0.3\protect \)
(\protect\( \Omega _{0}=1\protect \)) model, we use 5 (7) independent ray-tracing
realizations (see \protect\cite{jain}) to estimate the cosmic variance in a \protect\( 9\, \protect \)deg\protect\( ^{2}\protect \)
survey, which is then rescaled to the cosmic variance we should obtain for a
\protect\( 100\, \protect \)deg\protect\( ^{2}\protect \) survey. Given the
low number of realizations, the values here can only be used as a good estimation
of the order of magnitude of \protect\( \CosVar \left( \left\langle \kappa ^{2}\right\rangle \right) \protect \)
and \protect\( \CosVar \left( \left\langle (\vec{\nabla }\kappa )^{2}\right\rangle \right) \protect \).
It also seems, from these figures that the cosmic variance of \protect\( \left\langle (\vec{\nabla }\kappa )^{2}\right\rangle \protect \)
is more degraded by the difference in filtering beams than the other.}
\end{table}
 An estimation of \( {\mathcal{A}}^{\bo }_{2n} \) is then given by,

\begin{equation}
\label{A2nRes}
\frac{{\mathcal{A}}^{\bo }_{2n}}{\left\langle B_{(\theta )}b_{\bo (\theta ,\theta _{\galm })}\right\rangle ^{2}}\sim \left( \frac{2.12\%}{r}\right) ^{2}\frac{1}{\Sigma /100\textrm{ deg}^{2}}.
\end{equation}
 Since \( r_{\pabo } \) is very comparable to \( r \), we very roughly estimate
\( {\mathcal{A}}^{\pabo }_{2n} \)
\begin{equation}
\frac{{\mathcal{A}}^{\pabo }_{2n}}{\left\langle B_{(\theta )}b_{\pabo (\theta ,\theta _{\galm })}\right\rangle ^{2}}\sim \left( \frac{2.12\%}{r}\right) ^{2}\frac{1}{\Sigma /100\textrm{ deg}^{2}}.
\end{equation}
 The same considerations gives 
\begin{eqnarray}
\frac{{\mathcal{A}}^{\natural }_{2n}}{\left\langle B_{(\theta )}b_{\natural (\theta ,\theta _{\galm })}\right\rangle ^{2}} & = & \frac{(2.12\%)^{2}}{\Sigma /100\textrm{ deg}^{2}}.
\end{eqnarray}
 There is no \( r \) dependency here; the diagram cross-correlates \( \kappa _{\cmbm } \)
and \( \kappa _{\galm } \).

We can approximate the remaining \( \mathcal{A} \)-terms. They should be smaller
than the former. We have 
\begin{eqnarray}
{\mathcal{A}}^{\natural }_{3n} & \sim  & \frac{1}{r_{\natural }^{2}}\frac{(2.12\%\times 3.7\%)^{2}}{\Sigma /100\textrm{ deg}^{2}}\left\langle B_{(\theta )}b_{\natural (\theta ,\theta _{\galm })}\right\rangle ^{2}\nonumber \\
 & \ll  & {\mathcal{A}}^{\natural }_{2n}\nonumber 
\end{eqnarray}
 and 
\begin{eqnarray}
{\mathcal{A}}^{\natural }_{3c} & \sim  & \frac{(2.12\%\times 3.7\%)^{2}}{\Sigma /100\textrm{ deg}^{2}}\left\langle B_{(\theta )}b_{\natural (\theta ,\theta _{\galm })}\right\rangle ^{2}\nonumber \\
 & \ll  & {\mathcal{A}}^{\natural }_{2c}.\nonumber 
\end{eqnarray}
 Then, only the \( {\mathcal{A}}^{\natural }_{1} \) and \( {\mathcal{A}}^{\natural }_{2} \)
terms (boxed on Fig. \ref{diagrep}) contribute substantially to the cosmic
variance of \( \Cross _{\natural } \). Since \( {\mathcal{A}}^{\natural }_{1} \)
and \( {\mathcal{A}}^{\natural }_{2} \) are respectively the cosmic variance
of \( \left\langle \Delta E^{2}\right\rangle  \) (resp. \( \left\langle (\vec{\nabla }E)^{2}\right\rangle ) \)
and of \( \left\langle \kappa ^{2}\right\rangle  \) (resp. \( \left\langle (\vec{\nabla }\kappa )^{2}\right\rangle  \)),
we can write the variance of \( \Cross _{\natural } \) as 

\begin{eqnarray}
\CosVar (\Cross _{\bo }) & = & \\
 &  & \hspace {-2cm}\CosVar \left( \left\langle \Delta E^{2}\right\rangle \right) +\left( \frac{1+r^{2}}{2\, r^{2}}\right) \CosVar \left( \left\langle \kappa ^{2}\right\rangle \right) .\nonumber 
\end{eqnarray}
and
\begin{eqnarray}
\CosVar (\Cross _{\pabo }) & = & \\
 &  & \hspace {-2cm}\CosVar \left( \left\langle (\vec{\nabla }E)^{2}\right\rangle \right) +\left( \frac{1+r_{\pabo }^{2}}{2\, r_{\pabo }^{2}}\right) \CosVar \left( \left\langle (\vec{\nabla }\kappa )^{2}\right\rangle \right) .\nonumber 
\end{eqnarray}
Table \ref{CosVar.Bb} presents numerical results for various filtering scenarii
and models. 
\begin{table}
{\centering \begin{tabular}{ccccc}
&
\multicolumn{2}{c}{\( \CosVar \left( \Cross _{\bo }\right)  \) }&
\multicolumn{2}{c}{ \( \CosVar \left( \Cross _{\pabo }\right)  \)}\\
&
\( \Omega _{0}=0.3 \)&
\( \Omega _{0}=1 \)&
\( \Omega _{0}=0.3 \)&
\( \Omega _{0}=1 \)\\
\( \theta =5',\, \theta _{\galm }=2.5' \) &
6.44\%&
4.77\%&
6.06\%&
4.72\%\\
\( \theta =5',\, \theta _{\galm }=5' \) &
6.58\%&
4.79\%&
4.99\%&
4.23\%\\
\( \theta =10',\, \theta _{\galm }=5' \) &
8.71\%&
6.73\%&
9.49\%&
7.62\%\\
\end{tabular}\par}

\caption{\label{CosVar.Bb}Values of the cosmic variance of \protect\( \Cross _{\natural }\protect \).
The survey size is \protect\( 100\, \textrm{deg}^{2}\protect \). We used the
results presented in Table\ref{CosVar.Tab} and Fig. \ref{VarPol}. The \protect\( r_{\natural }\protect \)
parameters are assumed to be equal and set to 0.4. We didn't take into account
the filtering effects in the definition of \protect\( r\protect \). The difference
due to filtering correction is small, though. From this estimations, we can
expect a cosmic variance for \protect\( \Cross _{\natural }\protect \)of less
than 10\% for realistic scenarii.}
\end{table}

The two quantities, \( b_{\bo } \) and \( b_{\pabo } \), lead to similar cosmic
variance that are rather small. Obviously it would be even better to use \( b=b_{\bo }+b_{\pabo } \).
For such a quantity the resulting cosmic variance for the cross-correlation
coefficient should even be smaller, by a factor \( \sqrt{2} \), although a
detailed analysis is made complicated because of the complex correlation patterns
it contains.

\section{Conclusion}

We have computed a first order mapping that describes, in real space, the weak
lensing effects on the \comc polarization. In particular we derived the explicit
mathematical relation between the primary \comc polarization and the shear field
at leading order in lens effect. It demonstrates that a \( B \)-component of
the polarization field can be induced by lens couplings. We have shown however
that the \( B \)-map alone cannot lead to a non-ambiguous reconstruction of
the projected mass map.

Nonetheless, the \( B \)-component can potentially exhibit a significant correlation
signal with local weak lensing surveys. This opens a new window for detecting
lens effects on \comc maps. In particular, and contrary to previous studies
involving the temperature maps alone, we found that such a correlation can be
measured with a rather high signal to noise ratio even in surveys of rather
modest size and resolution. Anticipating data sets that should be available
in the near future, (\( 100\, \textrm{deg}^{2} \) survey, with \( 5' \) resolution
for galaxy survey and \( 10' \) Gaussian beam size for \comc polarization detection),
we have obtained a cosmic variance around \( 8\% \). Needless is to say that
this estimation does not take into account systematics and possible foreground
contaminations. It shows anyway that \Comc polarization contains a precious
window for studying the large scale mass distribution and consequently putting
new constraints on the cosmological parameters.

In this paper we have investigated specific quantities that would accessible
to observations. They both would permit to put constraint on the cosmological
constant. The simulated maps we presented here are only of illustrative interest.
We plan to complement this study with extensive numerical experiments to validate
our results (in particular on the cosmic variance), and explore the effect of
realistic ingredients we did not include in our simple analytical framework,
a shear non-gaussianity, lens-lens coupling and so forth.

\acknowledgements

We thank B. Jain, U. Seljak and S. White for the use of their ray-tracing simulations.
KB and FB thank CITA for hospitality and LvW is thankful to SPhT Saclay for
hospitality. We are all grateful to the TERAPIX data center located at IAP for
providing us computing facilities.

\end{document}